\pdfoutput=1
 \documentclass[aps,prd,reprint,twocolumn,showpacs,superscriptaddress,nofootinbib]{revtex4-1}  

\usepackage{tabularx}
\makeatletter
\def\hlinewd#1{%
\noalign{\ifnum0=`}\fi\hrule \@height #1 %
\futurelet\reserved@a\@xhline}
\makeatother
 \usepackage{auncial}
\usepackage{scrextend}
\usepackage{relsize}
\usepackage{amsmath}
\usepackage{amssymb}
\usepackage{epsfig}
\usepackage{graphicx}
\usepackage{hyperref}
\usepackage{dcolumn}   
\usepackage{tabu}
\usepackage{boldline}
\usepackage{slashed}
\usepackage{multirow}
\usepackage{bbold}
\usepackage{color}
\usepackage[normal]{subfigure}
\usepackage{rotating}
\usepackage[margin=0.9in,a4paper]{geometry}
\usepackage[table]{xcolor}
\usepackage{enumitem}
\usepackage[normalem]{ulem}
\usepackage{colortbl}
\usepackage{array,multirow}
\definecolor{nicered}{rgb}{0.7,0.1,0.1}
\definecolor{nicegreen}{rgb}{0.1,0.5,0.1}
\definecolor{red}{rgb}{1.0, 0, 0}

\allowdisplaybreaks

\definecolor{Grn}{rgb}{0.,0.75,0.}
\definecolor{Blu}{rgb}{0.,0.,1.}

\def\Im{\mbox{Im}\,}
\def\Re{\mbox{Re}\,}
\def\Tr{\mbox{Tr}\,}

\def\diag{\mbox{diag}\,}

\def\gsim{\raise0.3ex\hbox{$\;>$\kern-0.75em\raise-1.1ex\hbox{$\sim\;$}}}
\def\lsim{\raise0.3ex\hbox{$\;<$\kern-0.75em\raise-1.1ex\hbox{$\sim\;$}}}

\def\mb[#1]{\mathbf{#1}}

\renewcommand{\bar}{\overline}

\definecolor{LightCyan}{rgb}{0.88,1,1}
\definecolor{piggypink}{rgb}{0.99, 0.87, 0.9}
\definecolor{applegreen}{rgb}{0.55, 0.71, 0.0}
\definecolor{darkpastelgreen}{rgb}{0.01, 0.75, 0.24}
\definecolor{green-yellow}{rgb}{0.68, 1.0, 0.18}

\newcommand{\beq}{\begin{equation}}
\newcommand{\eeq}{\end{equation}}
\newcommand{\beqa}{\begin{eqnarray}}
\newcommand{\eeqa}{\end{eqnarray}}
\newcommand{\eps}{\epsilon}
\newcommand{\eV}{{\, \rm eV}}
\newcommand{\GeV}{{\, \rm GeV}}

\newcommand{\mueV}{{\, \mu {\rm eV}}}

\begin{document}

\begin{flushright}
{\footnotesize TTP-20-005}
\end{flushright}


\title{Quark Flavor Phenomenology of the QCD Axion}

\author{Jorge Martin Camalich}
\affiliation{Instituto de Astrof\'isica de Canarias, C/ V\'ia L\'actea, s/n
E38205 - La Laguna, Tenerife, Spain
}
\affiliation{%
Universidad de La Laguna, Departamento de Astrof\'isica, La Laguna, Tenerife, Spain
}
\author{Maxim Pospelov}
\affiliation{School of Physics and Astronomy, University of Minnesota, Minneapolis, MN 55455, USA}
\affiliation{William I. Fine Theoretical Physics Institute, School of Physics and Astronomy, University of Minnesota, Minneapolis, MN 55455, USA}
\author{Pham Ngoc Hoa Vuong}
\affiliation{Laboratoire de Physique Subatomique et de Cosmologie, Universit\' e Grenoble-Alpes, CNRS/IN2P3, Grenoble INP, 38000 Grenoble, France}

\author{Robert Ziegler}
\affiliation{Theoretical Physics Department, CERN, Geneva, Switzerland}
\affiliation{Institut f\"ur Theoretische Teilchenphysik (TTP),Karlsruher Institut f\"ur Technologie (KIT), 76131 Karlsruhe, Germany}

\author{Jure Zupan}
\affiliation{\normalsize \it 
Department of Physics, University of Cincinnati, Cincinnati, Ohio 45221,USA
}


\begin{abstract}
\noindent Axion models with generation-dependent Peccei-Quinn charges can lead to flavor-changing neutral currents, thus motivating QCD axion searches at precision flavor experiments. 
We rigorously derive limits on the most general effective flavor-violating couplings from current measurements and assess their discovery potential.  For two-body decays we use  available experimental data to derive limits on $q\to q' a$ decay rates for all flavor transitions. 
Axion contributions to neutral-meson mixing are calculated in a systematic way using chiral perturbation theory and  operator product expansion. We also discuss in detail baryonic decays and three-body meson decays, which can lead to the best search strategies for some of the couplings. For instance, a strong limit on the $\Lambda\to n a$ transition can be derived from the supernova SN~1987A. In the near future, dedicated searches for $q\to q' a$ decays 
at ongoing experiments could potentially test Peccei-Quinn  breaking scales up to $10^{12}$ GeV at NA62 or KOTO, and up to $10^{9}$  GeV at Belle II  or BES III. 
\end{abstract}

\maketitle


\section{Introduction} 
The QCD axion is arguably one of the best-motivated particles beyond the Standard Model (SM). Originally predicted by the Peccei-Quinn (PQ) solution to the strong CP problem~\cite{Peccei:1977hh,Peccei:1977ur,Wilczek:1977pj,Weinberg:1977ma}, i.e.,~the apparent absence of large CP violation in strong interactions~\cite{Baluni:1978rf,Crewther:1979pi,Pospelov:2005pr,Kim:2008hd,Baker:2006ts,Tanabashi:2018oca,Graner:2016ses,Afach:2015sja}, the axion is also an excellent cold Dark Matter (DM) candidate in large parts of the parameter space~\cite{AxionDM1,AxionDM2,AxionDM3}. 

The original electroweak axion models~\cite{Wilczek:1977pj,Weinberg:1977ma} were strongly disfavored by the non-observation of flavor-violating $K^+\to \pi^+ a$ decays (where $a$ is the axion)~\cite{Goldman:1977en}. Since then, experimental searches for axions have been primarily focused on the flavor-conserving axion couplings. Searches using haloscopes~\cite{ADMX1, ADMX2, ADMXfuture,Brubaker:2016ktl,McAllister:2017lkb,TheMADMAXWorkingGroup:2016hpc,Majorovits:2016yvk,Brun:2019lyf} and helioscopes~\cite{CAST, IAXO1, BabyIAXO,Aprile:2014eoa} rely on axion couplings to photons, while searches using precision magnetometry rely on axion couplings to gluons~\cite{Budker:2013hfa}. Much activity is being devoted to devise new experimental techniques, and ambitious projects have been proposed for the upcoming years (for a review see Ref.~\cite{Graham:2015ouw}). 

In this paper we reiterate the importance of flavor violating transitions for axion searches. Indeed, a QCD axion with flavor violating couplings could well be first discovered in flavor-physics experiments. While this possibility was already contemplated in the literature for rare meson decays and neutral meson mixing~\cite{Wilczek:1982rv,Feng:1997tn, Kamenik:2011vy, Bjorkeroth:2018dzu}, we go well beyond the state of the art. We show that, in light of upcoming experiments, there are a number of additional dedicated searches that could and should be performed. Besides the two-body meson decays, also the three-body and baryonic decays can provide the best sensitivity to specific axion couplings. We provide a careful and rigorous analysis of the resulting constraints by exploiting the entire set of current experimental information. We also improve the predictions for axion induced neutral-meson oscillations by using effective-field theory methods, and derive a new bound from supernova cooling.

Predictions for axion couplings to fermions are model-dependent. The only requirement for  a successful axion model is an almost exact global $U(1)$ PQ symmetry that is spontaneously broken and anomalous with respect to QCD. Therefore, the only coupling shared by all axion models is the axion coupling to the CP-odd gluon operator, $G\tilde G$, arising from the QCD anomaly and responsible for the solution to the strong CP problem. Axion models divide into two classes, depending on how the color anomaly arises. In the KSVZ-type models~\cite{Kim:1979if, Shifman:1979if} the color anomaly is due to a set of heavy fermions that are vectorlike under the SM but chiral under the PQ symmetry. In this case the axion does not couple to elementary SM fermions at tree level. In the DFSZ-type models~\cite{Dine:1981rt, Zhitnitsky:1980tq}, on the other hand, the SM fermions carry PQ charges, and the axion always couples to the fermionic currents. Whether  these couplings are flavor conserving or flavor violating is a model-dependent choice.

In the original DFSZ model the PQ charges are taken to be flavor universal~\cite{Dine:1981rt, Zhitnitsky:1980tq}, so that flavor-violating axion couplings arise only at loop level. In general, though, flavor violating axion couplings are present already at tree level. This is  the case, for instance, in generalized DFSZ-type models with generation dependent PQ charges~\cite{Davidson:1984ik, Peccei:1986pn, Krauss:1986wx, Geng:1988nc, Celis:2014iua}, which can also allow to suppress the axion couplings to nucleons~\cite{DiLuzio:2017ogq, Bjorkeroth:2019jtx, Saikawa:2019lng}. Particularly motivated scenarios, which lead to flavor violating axion couplings at tree level, arise when the PQ symmetry is part of a flavor group that shapes the structure of the yukawa  sector~\cite{Davidson:1981zd, Wilczek:1982rv}. The PQ symmetry could enforce texture zeros in the Yukawa matrices~\cite{Davidson:1983fy, Davidson:1983tp, Bjorkeroth:2018ipq}, or be responsible for their hierarchical structure \`a la Froggatt-Nielsen (FN)~\cite{Froggatt:1978nt}. While in the simplest scenario PQ and FN symmetries are identified~\cite{Calibbi:2016hwq, Ema:2016ops, Ema:2018abj}, PQ could also be a subgroup of a larger flavor symmetry, see e.g. Refs.~\cite{Ahn:2014gva, Nomura:2016nfi, Bjorkeroth:2017tsz, Ahn:2018nfb, Ahn:2018cau, Linster:2018avp, Carone:2019lfc}.  Finally, flavored PQ symmetries can arise also in the context of Minimal Flavor Violation~\cite{Albrecht:2010xh, Arias-Aragon:2017eww} or as accidental symmetries in models with gauged flavor symmetries~\cite{Babu:1992cu, Cheung:2010hk, Suematsu:2018hbu, Bonnefoy:2019lsn}.

In our analysis we remain, for the most part, agnostic about the origin of the flavor and chiral structure of axion couplings to SM fermions, and simply treat axion couplings to fermions as  independent parameters in an effective Lagrangain. For related studies of axion-like particles with flavor violating couplings, see \cite{Cornella:2019uxs,Bauer:2019gfk,Albrecht:2019zul} (for loop induced transitions see \cite{Flacke:2016szy,Frere:1981cc,Freytsis:2009ct,Dolan:2014ska,Batell:2009jf,Gavela:2019wzg,Izaguirre:2016dfi,Dobrich:2018jyi}). We restrict the analysis to the case of the (practically) massless QCD axion, but our results can be repurposed for any other light scalar or pseudoscalar with flavor violating couplings to the SM fermions, as long as the mass of the (pseudo-)scalar is much smaller than the typical energy release in the flavor transition.

The paper is structured as follows. In Section \ref{sec:axion:couplings} we introduce our notation for the axion couplings to fermions and comment on their  flavor structure. In Section \ref{sec:bounds:hadron:decays} we derive the bounds on these couplings from two-body and three-body meson decays, from baryon decays and from baryon transitions in supernovae. Section \ref{sec:meson:mix} contains bounds from mixing of neutral mesons, Section~\ref{sec:diag} reviews bounds on flavor-diagonal couplings, and Section~\ref{sec:top:quark} discusses axion couplings involving the top quark. Finally, in Section \ref{sec:results} we present the results and experimental projections. Details about renormalization of effective axion couplings, experimental recasts of two-body meson decays and hadronic inputs are deferred to the Appendix.

\section{Axion Couplings to Fermions}
\label{sec:axion:couplings}

The Lagrangian describing the most general interactions of the axion with the SM fermions is given by~\footnote{Note that diagonal vector couplings are unphysical up to electroweak anomaly terms, which are irrelevant for the purpose of this paper.}
(see also Appendix \ref{sec:app:RGEs})
\begin{align}
\label{eq:couplings}
& \mathcal{L}_{aff} = 
\frac{\partial_\mu a}{2 f_a} \, \bar f_i \gamma^\mu \big( c^V_{f_i f_j} + c^A_{f_i f_j} \gamma_5 \big) f_j  \, , 
\end{align}
where $f_a$ is the axion decay constant, $c^{V,A}_{f_i f_j}$ are hermitian matrices in flavor space, and the sum over repeated generational indices, $i,j=1,2,3$, is implied. For future convenience we define effective decay constants as 
\begin{align}
\label{eq:FVA:def}
F^{V,A}_{f_ i f_j} & \equiv \frac{2 f_a}{c^{V,A}_{f_i f_j}} \, .
\end{align}
In general $F^{V,A}_{f_ i f_j}, i\ne j$, are complex, with $\big(F^{V,A}_{f_ i f_j}\big)^*=F^{V,A}_{f_ j f_i}$. Throughout the paper we take $a$ to be the QCD axion, so that  its mass is inversely proportional to $f_a$~\cite{Gorghetto:2018ocs}, 
\begin{align}
m_a = 5.691(51) \mueV \left( \frac{10^{12} \GeV}{f_a} \right) \, .
\end{align}
For the ``invisible'' axion the decay constant is $f_a \gg 10^6 \GeV$~\cite{Georgi:1986df}, in which case the axion is much lighter than an eV and essentially decoupled from the SM. We will always be working in this limit, so that in the flavor transitions the axion can be taken as massless for all practical purposes.

In this mass range the axion has a lifetime that is larger than the age of the universe, and therefore is a suitable DM candidate. If the PQ symmetry is broken before inflation, axions are produced near the QCD phase transition and yield the observed DM abundance for axion decay constants of the order $f_a \sim (10^{11} \div 10^{13}) \GeV$~\cite{AxionDM1,AxionDM2,AxionDM3}, assuming natural values of the misalignment angle. Other production mechanisms, e.g., via parametric resonance, allow for axion DM also for smaller decay constants, down to $f_a \sim 10^8 \GeV$~\cite{Co:2017mop}. We will see below that precision flavor experiments are able to test this most interesting region of the QCD axion parameter space. 

The axion couplings to the SM fermions  in the mass basis, $c^V_{f_i f_j}$ and $c^A_{f_i f_j}$, are related to the PQ charge matrices in the flavor basis, $X_f$, through
\begin{align}
c_{f_i f_j}^{V,A} & = \frac{1}{2N} \left( V_{f_R}^\dagger X_{f_R} V_{f_R} \pm V_{f_L}^\dagger X_{q_L} V_{f_L} \right)_{ij}  \, ,
\label{Cdef}
\end{align}
where $N$ is the QCD anomaly coefficient of the PQ symmetry. The unitary rotations $V_{f_L,f_R}$ diagonalize the  appropriate SM fermion yukawa matrices, $V_{f_L}^\dagger y_f V_{f_R}= y_f^{\rm diag}$, for  the ``up'' and ``down'' quark flavors, $f = u,d$.
We focus on axion couplings to quarks, and refer the reader to Ref.~\cite{LFVaxion} for present and future prospects for testing lepton flavor violating axion couplings. 
 Off-diagonal couplings arise whenever PQ charges, $X_{q_L}, X_{f_R}$, are not diagonal in the same basis as the yukawa matrices, $y_f$.
Their sizes depend on the misalignment between the two bases, parametrized by the unitary rotations $V_{f_L},V_{f_R}$ (taking $X_{q_L},X_{f_R}$ to be diagonal).

Very different flavor textures of $c_{f_i f_j}^{V,A}$ are possible. Provided a suitable set of PQ charges and appropriate flavor structures of the SM yukawa matrices, it is possible for just a single off-diagonal coupling to be large $c^{V,A}_{f_if_j}\sim {\cal O}(1), i\ne j,$ with all the other off-diagonal couplings zero or very small.  For example, one can realize a situation where $c^V_{bs} = c^A_{bs} \sim 1$ and all the other $c^{V,A}_{f_i \ne f_j} = 0$, by taking $X_{q_L} = X_{u_R} = \mathbb{1}$, $(X_{d_R})_2 \ne (X_{d_R})_3$, with down yukawa matrix $y_d$ such that the only non-vanishing rotation is in the 2-3 RH sector, $s_{23}^{Rd} \sim 1$. Moreover, while one would generically expect axial couplings $c^A_{f_i f_j}$ and vector couplings $c^V_{f_i f_j}$ to be of the same order, the latter can be suppressed in a situation where $X_{f_R} = - X_{q_L}$ and $V_{f_R} = V_{f_L}$, which can arise in models where PQ charges are compatible with a grand unified structure (see Ref.~\cite{Ernst:2018bib} for a recent example in SO(10)), and yukawas are hermitian, positive definite matrices (see e.g. Ref.~\cite{Moorhouse:2008yd} for a realization of this scenario in SO(10)). 

In the absence of a theory of flavor, we will be agnostic about the origin of the possible flavor misalignment, and simply take $c_{f_i f_j}^{V,A}$ to be unknown parameters in an effective Lagrangian, which will be constrained solely from data.


\begin{table*}[t]
\centering
\begin{tabular}{lcccc}
\hline\hline\\[-3mm]
Decay &   $sd$ & $cu$ & $bd$ & $bs$ \\
\hline\\[-3mm]
${\rm BR} (P_1 \to P_2 + a)$ &
$7.3 \times 10^{-11}$~\cite{Adler:2008zza} & 
no analysis &	
$4.9 \times 10^{-5}$~\cite{Ammar:2001gi}  &
$4.9 \times 10^{-5}$~\cite{Ammar:2001gi} \\
${\rm BR} (P_1 \to P_2 + a)_{\rm recast}$ & no need
& $8.0 \times 10^{-6}$~\cite{CLEO2008}
&  $2.3 \times 10^{-5}$~\cite{Aubert:2004ws}
&
$7.1 \times 10^{-6}$~\cite{BABAR13}   \\
${\rm BR} (P_1 \to P_2 + \nu \overline{\nu})$ &
 $1.47^{+ 1.30}_{- 0.89} \times 10^{-10}$~\cite{Adler:2008zza}  & no analysis 
 & 
 $0.8 \times 10^{-5}$~\cite{BELLE17}  & 
 $1.6 \times 10^{-5}$~\cite{BELLE17} \\
 \hline
 \hline\\[-3mm]
 ${\rm BR} (P_1 \to V_2 + a)$ &
  $3.8 \times 10^{-5}$~\cite{Adler:2000ic}  
 & no analysis & no analysis  & no analysis \\
 ${\rm BR} (P_1 \to V_2 + a)_{\rm recast}$ & no need
  & no data
 & no data 
 & $5.3 \times 10^{-5}$~\cite{BABAR13} \\
 ${\rm BR} (P_1  \to V_2 + \nu \overline{\nu})$ & 
$4.3 \times 10^{-5}$~\cite{Adler:2000ic}   
& no analysis 
 &
 $2.8 \times 10^{-5}$~\cite{BELLE17}  & $2.7 \times 10^{-5}$~\cite{BELLE17} \\
 \hline\hline
  \end{tabular}
  \caption{Experimental inputs for meson decays, see text for details. We show the 90\% CL upper bounds on the branching ratios of a pseudo-scalar meson $P_1$ to  another pseudo-scalar ($P_2$) or vector ($V_2$) meson (for $sd$ transitions $V_2 = \pi \pi$ instead). The bounds shown are for decays to neutrinos or massless invisible axions. In the latter case we also show 
  our bounds obtained by recasting related searches for invisible decays (subscript "recast").
   \label{tab:ExpInput} }
\end{table*}

\section{Bounds from hadron decays}

\label{sec:bounds:hadron:decays}
Bounds on the vector and axial-vector parts of the flavor-violating axion couplings, Eqs.~\eqref{eq:couplings}, \eqref{eq:FVA:def},  can be derived from searches for hadron decays with missing energy. 
In this section we consider the bounds from two-body decays of pseudoscalar mesons to pseudoscalar and vector mesons respectively, from three-body meson decays, and from decays of baryons. In each case, we first review the available and planned experimental measurements  and then interpret them in terms of the bounds on flavor violating axion couplings. 

\subsection{Bounds from two-body meson decays}
\label{sec:bounds:2bodymeson:decays}

Due to parity conservation of strong interactions, the $P_1\to P_2 a$ decays of a pseudoscalar meson $P_1$ to a pseudoscalar meson $P_2$ are only sensitive to the vector couplings of the axion, while the $P_1\to V_2 a$ decays, where $V_2$ is the vector meson, are only sensitive to the axial-vector couplings (see Appendix~\ref{sec:app:hadronic}). Searches targeting specifically the massless axion were performed in $K^+\to\pi^+ a$~\cite{Adler:2008zza}, $B^+\to K^+ a$~\cite{Ammar:2001gi} and $B^+\to \pi^+ a$~\cite{Ammar:2001gi} decays. In addition, searches for SM decays where the invisible final state is a $\nu\bar\nu$ pair can be recast to derive limits on the axion couplings. This requires that the two-body kinematics of an (essentially) massless axion is included in the search region, as was the case in the BaBar and CLEO searches for $B \to K^{(*)} \nu\bar\nu$~\cite{BABAR13},  $B \to \pi \nu\bar\nu$~\cite{Aubert:2004ws} and $ D\to(\tau \to\pi \bar\nu)\nu$~\cite{CLEO2008}.   
Note that the corresponding Belle data sets analyzed in Refs.~\cite{Lutz:2013ftz,BELLE17} cannot be readily used to set bounds on axion couplings, because the analyses either cut out two-body decays with a massless axion~\cite{Lutz:2013ftz} or used multi-variate methods~\cite{BELLE17}  that are difficult to re-interpret for different purposes~\cite{HeckTalk} (see also Ref.~\cite{Filimonova:2019tuy}). 

The available experimental information is summarized in 
Table~\ref{tab:ExpInput}, where we give the 90\% CL upper limits on the branching ratios for decays involving neutrinos or invisible massless axions. We include the limits on the decays involving axions that we obtained by recasting the experimental searches for decays involving neutrinos (``recast"). Table~\ref{tab:ExpInput}  
shows bounds on the branching ratios for decays to pseudoscalar mesons, $P_1\to P_2 a$, for $K^+\to \pi^+ a$ ($s\to d$ transition, experimental analysis in Ref.~\cite{Adler:2008zza}),  $D^+\to \pi^+a$ ($c\to u$ transition, our recast of Ref.~\cite{CLEO2008}), $B^+\to \pi^+ a$ ($b\to d$ transition, experimental analysis in Ref.~\cite{Ammar:2001gi} and our recast of Ref.~\cite{Aubert:2004ws}), $B^+\to K^+ a$ ($b\to s$ transition, experimental analysis in Ref.~\cite{Ammar:2001gi} and our recast of Ref.~\cite{BABAR13}). For decays to vector mesons, $P_1\to V_2 a$, there is experimental information on the $B\to K^* a$ decay ($b\to s$ transition, our recast of Ref.~\cite{BABAR13}).  In the same section of Table~\ref{tab:ExpInput} we also include the bounds on  $K^+\to \pi^+ \pi^0 a$ decay used in Sec.~\ref{sec:3body} below  ($s\to d$ transition, experimental analysis in Ref.~\cite{Adler:2000ic}). For details on the recast see Appendix~\ref{sec:app:recast}.

The above analyses could be improved with dedicated axion searches applied to existing data or to ongoing experiments. Sensitivity to the $K^+\to \pi^+ a$ decay better than the present world best limit can be achieved at the NA62 experiment. For a massless axion, an improvement by an order of magnitude  compared to the BNL result, ${\rm BR}(K^+\to \pi^+ a)<7.3 \times 10^{-11}$~\cite{Adler:2008zza}, is expected  by 2025~\cite{Ruggieroprivate}. The same flavor transition is probed by KOTO searching for the neutral decay mode $K_L\to \pi^0 a$. The current limit using data collected in 2015 is ${\rm BR}(K_L\to \pi^0 a)<2\times10^{-9}$  and is in the same ballpark as for the decay to the neutrino pair~\cite{Ahn:2018mvc}. The KOTO collaboration expects the sensitivity
  to be improved down to the $10^{-11}$ level~\cite{NanjoTalk}.
Improved sensitivity in the neutral-kaon mode can also be expected from the proposed KLEVER experiment~\cite{Ambrosino:2019qvz}. 

To the best of our knowledge, the future prospects for the heavy-meson axion decays $P_1\to P_2 a$ or $P_1\to V_2 a$ have not yet been estimated by the experimental collaborations. Nevertheless, it is clear that improvements over the present situation are justifiably expected. 
For instance, the experimental error on the $D^+\to \tau^+\nu$ branching ratio is projected to be reduced by a factor of 3 with $20 \, {\rm fb}^{-1}$ integrated luminosity at BESIII~\cite{Ablikim:2019hff} compared to the present value~\cite{Ablikim:2019rpl}, indicating a potential significant improvement in sensitivity to the $c\to u a$ transition. 
The reach on the branching ratios for the axionic decay modes of $B$ mesons will be improved with the amount of data expected to be collected at Belle II, which is roughly a factor of $50$ larger compared to the BaBar and Belle samples. 

 Several potentially interesting channels are lacking any experimental analyses so far. For example, there is no  experimental analysis of  $c\to u a$ transitions that are sensitive to the axial-vector coupling, i.e.,  there are no $D \to \pi \pi X_{\rm inv}$ or $D \to \rho X_{\rm inv}$,  $X_{\rm inv}=\nu \overline{\nu},a$, searches. One could also search for a $c\to u a$ signal in $D_s\to K a$, $D_s\to K^* a$ decays, all of which could be performed at Belle II and BESIII. 
 Potentially, LHCb could also probe these couplings using decay chains, such as $B^-\to D^0\pi^-$ followed by $D^0\to \rho^0a$, which results in three charged pions + MET and two displaced vertices.  The lack of such analyses means that there is at present no bound from meson decays on axial $cu$ couplings to the axion. Similarly, there is at present no publicly available experimental analysis that bounds the $B \to \rho a$ decays (as discussed above, one cannot readily use  for that purpose the $B \to \rho \nu\bar \nu$ Belle data from Ref.~\cite{BELLE17}, while BaBar has not performed such an analysis). Finally, our recast bounds  on $B \to K^{(*)}a, B \to \pi a$ could be easily improved by dedicated experimental searches using already collected data. At LHCb one could measure the $B\to K^* a$ and $B\to \rho a$ branching ratios using the decay chains such as $\bar B_s^{0**}\to K^+ B^-$ or $\bar B^{0**}\to \pi^+ B^-$ followed by  $B^-\to K^{*-}(\to K_S \pi^-)a $, or $\bar B_s^{0**}\to K_S \bar B^0$ followed by $\bar B^0 \to \bar K^{*0} a, \rho^0 a$ \cite{Stone:2014mza}.
 One could also attempt
 more challenging decay chain measurements such as $B_s^*\to B_s\gamma$, followed by $B_s\to \phi a$ or $B_s\to K^* a$.  

We now convert the bounds on the branching ratios in Table~\ref{tab:ExpInput} to bounds on flavor violating couplings of axions to quarks, Eqs. \eqref{eq:couplings}, \eqref{eq:FVA:def}. 
The corresponding partial decay widths
 are given by
\begin{align}
\Gamma =\kappa_{12} 
\begin{cases} 
\frac{f_+(0)^2}{|F^{V}_{ij}|^2} \, , & P_1 \to P_2 a  
\\
\frac{A_0(0)^2}{|F^{A}_{ij}|^2} \, , &  P_1 \to V_2 a
\end{cases}
\end{align}
with the kinematic prefactor
\beq
\label{eq:kappa12}
\kappa_{12}=\frac{M_1^3}{16\pi} \left(1-\frac{M_2^2}{M_1^2}\right)^3,
\eeq
where $M_1$ ($M_2$) is the mass of the parent (daughter) meson. Since $K_L\to \pi^0a$ decay is CP violating, the partial decay width in that case is given by 
\beq
\Gamma_{K_L\to \pi^0 a}=\kappa_{12} f_+(0)^2\big[\Im (1/F^{V}_{sd})\big]^2,
\eeq 
and thus vanishes in the CP conserving limit, $\Im F^{V}_{sd}=0$, cf. Eq.~\eqref{eq:FVA:def}. The $K_L\to \pi^0 a$ and $K^+\to \pi^+ a$ decay rates obey the Grossman-Nir bound ${\rm BR}(K_L\to \pi^0 a)\leq 4.3 \,  {\rm BR}(K^+\to \pi^+a)$~\cite{Grossman:1997sk,Kitahara:2019lws}. 

The form factors $f_+(q^2)$ and $A_0(q^2)$ are defined in Appendix~\ref{sec:app:hadronic}, where we also collect the numerical values used as inputs in the numerical analysis. The resulting bounds on axion couplings $F^{V,A}_{ij}$ are shown in Tab. \ref{tab:main:results}. The implications of these results and future projections will be discussed in Sec.~\ref{sec:results}.

\subsection{Bounds from three-body meson decays}\label{sec:3body}

The E787 experiment at Brookhaven performed a search for the three-body $K^+\to\pi^0\pi^+ a$ decay mediated by the $s\to d a$ transition, 
and set the bound $\text{BR}(K^+\to\pi^0\pi^+ a)\leq 3.8 \times 10^{-5}$ at 90\% CL~\cite{Adler:2000ic}. The related decay mode $K_L\to\pi^0\pi^0 a$ has also been searched for, resulting in the upper limit for light massive axions  $\text{BR}(K_L\to\pi^0\pi^0 a)\lesssim0.7\times 10^{-6}$~\cite{E391a:2011aa}. However, this analysis excluded the $m_a=0$ kinematic region and is thus not applicable to the case of the QCD axion~\cite{Suzukiprivate}. Other decay modes such as $K_L\to \pi^+\pi^- a$ or those involving the decays of $K_S$ have not been investigated experimentally.   

Parity conservation implies that the  $K\to\pi\pi a$ decays are sensitive only to the axial-vector couplings of the axion to quarks (see Appendix~\ref{sec:app:hadronic}). The form factors entering the predictions are related via isospin symmetry to the form factors measured in $K^+\to\pi^+\pi^- e^+\nu$~\cite{Batley:2010zza,Batley:2012rf,Littenberg:1995zy,Chiang:2000bg}, making  precise predictions for $K\to \pi\pi a$ decay rates possible. The two final state pions can be only in the total isospin $I=0$ or the $I=1$ state, since the $s\to d a$ Lagrangian is $|\Delta I|=1/2$, while the initial kaon is part of an isodoublet. Bose symmetry demands the decay amplitude to be symmetric with respect to the exchange of the two pions. The $I=0$ ($I=1$) amplitude is even (odd) under this permutation. The form factors must therefore enter in combinations which are even (odd) with respect to the exchange of pion momenta, $p_{\pi_1} \leftrightarrow p_{\pi_2}$. The two pions in the decay $K^0\to\pi^0\pi^0 a$ ($K^+\to\pi^+\pi^0 a$) are in a pure $I=0$ ($I=1$) state and one obtains,
\beq\label{eq:Kpipi:KL}
\frac{d\Gamma(K_L\to\pi^0\pi^0 a)}{ds}=\big[\Re(1/F_{sd}^A)\big]^2\frac{(m_{K^0}^2-s)^3}{1024\pi^3m_K^5}\beta F_s^2,
\eeq
and
\beq\label{eq:Kpipi:Kp}
\begin{split}
\frac{d\Gamma(K^+\to\pi^+\pi^0 a)}{ds}&=\frac{1}{|F_{sd}^A|^2}\frac{(m_{K^+}^2-s)^3}{1536\pi^3m_K^5}\beta\\
&\times\left(F_p^2+\beta^2G_p^2+2\beta F_p G_p\right),
\end{split}
\eeq
where $s=(p_{\pi_1}+p_{\pi_2})^2$ and $\beta^2=1-4m_\pi^2/s$. The functions $F_s$, $F_p$ and $G_p$ correspond to the moduli of the coefficients in the partial-wave expansion of the form factors (which are complex functions of the kinematic variables of the two-pion system) in the $K^+\to\pi^+\pi^- e^+\nu$ decay channel~\cite{Pais:1968zza,Bijnens:1994ie,Batley:2010zza,Batley:2012rf} (see Appendix~\ref{sec:app:hadronic}). In the first equation we also used the fact that the final state is CP-odd and, therefore, the decay occurs through the predominantly CP-odd component of $K_L$. 

One can also obtain expressions for other decay modes from eqs.~\eqref{eq:Kpipi:KL} and \eqref{eq:Kpipi:Kp}. The amplitude of $K_L\to\pi^+\pi^- a$ contains both $I=0$ and $I=1$ isospin components and the decay rate in the isospin limit is,
\begin{align}
 \Gamma(K_L\to\pi^+\pi^- a)&=\tilde \Gamma(K^+\to\pi^+\pi^0 a)\nonumber\\
 &+2 \, \Gamma(K_L\to\pi^0\pi^0 a), 
\end{align}
where $\tilde \Gamma(K^+\to\pi^+\pi^0 a)$ is obtained by replacing $1/|F_{sd}^A|\to \Re(1/F_{sd}^A)$ in $\Gamma(K^+\to\pi^+\pi^0 a)$. Rates for the $K_S$ decays are obtained by replacing  $\Re(1/F_{sd}^A)\to\Im(1/F_{sd}^A)$ in the rates of $K_L$.

Better sensitivity to the axial $sd$ axion coupling can be achieved by searching for $K_L\to \pi^0\pi^0a$ at KOTO~\cite{Suzukiprivate}, while a search for $K^+\to \pi^+\pi^0a$ could be attempted at NA62. Note that eqs.~\eqref{eq:Kpipi:KL} and \eqref{eq:Kpipi:Kp} can be combined to the inequality
\begin{align}
{\rm BR}(K_L\to\pi^0\pi^0 a)\leq 31\, {\rm BR}(K^+\to\pi^+\pi^0 a),
\end{align}
which, besides the ratio of kaon lifetimes $\tau_L/\tau_+=4.1$ (that gives the main effect in the Grossman-Nir bound for ${\rm BR}(K\to \pi a)$), also includes the effects of different combinations of Clebsch-Gordan coefficients, form factors and phase space factors. 

The numerical input values for the hadronic matrix elements are described in Appendix~\ref{sec:app:hadronic}, while the resulting bound on the axion coupling obtained from the current bound on $K^+\to\pi^+\pi^0 a$~\cite{Adler:2000ic}, is included in Tab. \ref{tab:main:results}. The implications of these results and prospects are discussed further in Sec.~\ref{sec:results}. 

Other searches using three body decays could be of interest.  Decays such as $B^+\to \rho^0\pi^+ a$, $B^0\to \rho^0 \rho^0 a$ or $B_s\to K_S \rho^0 a$, $B_s\to K^{*0}\rho^0 a$, $B_s\to \phi \rho^0 a$ could potentially be even attempted at LHCb, since they result in three or four charged particles and a massless invisible particle. Other possible decays of theoretical interest, but probably only measurable at Belle II, are  $B^+\to \pi^0\pi^+ a$, $B^0\to \pi^+\pi^- a$, $B^0\to \rho^+\rho^- a$,  $B^+ \to \rho^+\rho^0 a$. Belle II could also access from the $\Upsilon(5S)$ run  the $B_s$ decays with neutral pions such as  $B_s\to K_S \pi^0 a$, $B_s\to K^{*0}\pi^0 a$, etc. Providing predictions for these decays lies beyond the scope of the present paper, though controlled calculations using QCD factorization and soft-collinear effective theory may be possible \cite{Klein:2017xti} and could be attempted in the future. Until then one can use as a rough guide the bounds that were obtained for the related two body decays, i.e., $B^+\to \pi^+ a$ (on $F_{bd}^V$) and $B^+\to \rho^+ a$ (on $F_{bd}^A$), as a ball-park figure for what an interesting experimental reach for $B^+\to \rho^0\pi^+ a$ (bounding a combination of $F_{bd}^V$ and $F_{bd}^A$) might be. 

Finally one can also use LHCb di-muon data collected for the $B_{q} \to \mu \mu$ analysis to constrain the axial  couplings $F^A_{bd}$ and $F^A_{bs}$. As long as no vetos on extra particles in the event are applied in the LHCb analysis, their datasets can be used to constrain decays with additional particles in the final state such as axions. In Ref.~\cite{Albrecht:2019zul} the present data were used to derive constraints on $F^A_{bq}$ of the order of $10^{5} \GeV$, cf. Table~\ref{tab:main:results}. With 300 fb$^{-1}$ the bounds can be strengthened by about an order of magnitude, and could be further improved by using also the ATLAS and CMS data on $B_{q} \to \mu \mu$. The same strategy might also be applied to $sd$ transitions using $K_S \to \mu \mu$ decays, as proposed in Ref.~\cite{Junior:2018odx}. 

\subsection{Bounds from Baryon Decays}

Baryon decays, $B_1\to B_2 a$, are sensitive to both axial and vector couplings of the axion. The 
decay rates are given by
\beq
\Gamma(B_1\to B_2 a)=\kappa_{12} \bigg(\frac{f_1(0)^2}{|F^{V}_{ij}|^2}+\frac{g_1(0)^2}{|F^{A}_{ij}|^2}\bigg),
\eeq
with the kinematic prefactor $\kappa_{12}$ given in Eq.~\eqref{eq:kappa12}. The form factors $f_1(q^2)$ and $g_1(q^2)$ are discussed in detail in  Appendix~\ref{sec:app:hadronic}.

At present there are no published experimental searches for $B_1\to B_2 a$ decays. We therefore set 90\% CL upper limits on  ${\rm BR}(B_1\to B_2 a)$ indirectly. For hyperons, we subtract from unity the branching ratios for all  channels that have been measured so far, adding the experimental errors in quadrature~\cite{Tanabashi:2018oca}.
For $\Lambda_b$ decays we use the SM prediction for its lifetime at NLO in $\alpha_s$ and at ${\mathcal O}(1/m_b)$ in heavy quark expansion ~\cite{Lenz:2014jha}, compare it to the experimental measurements~\cite{Tanabashi:2018oca}, and ascribe the difference to the allowed value for ${\rm BR}(\Lambda_b \to B_2 a)$. For $\Lambda_c$ we saturate the observed lifetime with the $\Lambda_c^+\to p a$ decay width. The resulting upper bounds on the branching ratios ${\rm BR}(B_1\to B_2 a)$ are collected in Tab.~\ref{tab:widthlimits} and used to  derive the limits on axion couplings $F^{V,A}_{ij}$ shown in Tab. \ref{tab:main:results}. 

The sensitivity could be improved substantially with dedicated searches for $B_1\to B_2 a$ decays. The BESIII collaboration plans to measure $B_1\to B_2 \nu\bar\nu$ decays of hyperons by using a sample of hyperon-antihyperon pairs collected at the $e^+ e^- \to J/\psi$ peak~\cite{Li:2016tlt}.
At BESIII one could also study the decays of charmed baryons into axions. 
Namely, approximately $\sim 10^4$ $\Lambda_c^+ \overline{\Lambda}_c^-$ pairs have been collected in a run with $567$ pb$^{-1}$ of integrated luminosity at $e^+e^-$ collisions just above the pair-production threshold~\cite{Ablikim:2015prg}. 
Note that bottom baryons are not produced in the modern $e^+e^-$ machines, since they run at energies below the corresponding pair-production thresholds, so that only LHCb, while challenging, could have access to these decays.

\begin{table}[t]
\centering
\begin{tabular}{cc}
\hline\hline\\[-3mm]
Baryon $B_1$ & ${\rm BR}(B_1 \to B_2 a)_{\rm 90\%}$ \\
\hline\\ [-3mm]
$\Lambda$ & $8.5\times 10^{-3}$ \\
$\Sigma^+$ & $4.9 \times 10^{-3}$\\
$\Xi^0$ & $2.3\times 10^{-4}$\\
$\Xi^-$ & $6.4\times 10^{-4}$\\
\hline\\[-3mm]
$\Lambda_c$&1\\
$\Lambda_b$& $4.1\times 10^{-2}$\\
\hline\hline
\end{tabular}
  \caption{The 90\% CL upper bounds on the  branching fractions for the baryon $B_1 \to B_2 a$ decays obtained 
  by adding up the measured branching fractions of the exclusive modes (hyperons) or by comparing theory predictions for lifetimes with the measurements (heavy baryons). 
  \label{tab:widthlimits}}
\end{table}

\subsection{Supernova bound}
\label{sec:bounds:SN}

In the core of neutron stars (NS) hyperons coexist in equilibrium with neutrons, protons and electrons~\cite{1960SvA,Prakash:1996xs,Lattimer:2006xb,Vidana:2018bdi}. The decay $\Lambda\to n a$ would represent a new cooling mechanism for NS, and can thus be constrained by stellar structure calculations and observations. At exactly zero temperature, the degenerate $\Lambda$ and neutron distributions must have the same Fermi energy, leaving no phase space for the $\Lambda\to n a$ decays to occur. The degeneracy is partially lifted at finite temperature allowing for the $\Lambda\to n a$ transitions with a rate that increases with the temperature. The impact of this new cooling mechanism is maximal during the few seconds after the supernova explosion, when a proto-neutron star (PNS) reaches temperatures of several tens of MeV~\cite{Burrows:1986me,Bethe:1990mw}. 

In order to estimate the cooling facilitated by the $sd$-axion interaction in this early phase of the supernova evolution we assume that the PNS is a system of non-interacting (finite temperature) Fermi gases of neutrons, protons, electrons and $\Lambda$ baryons that are in thermal and chemical equilibrium. Furthermore, we assume that the neutrinos are trapped inside the PNS, while the lepton fraction number, relative to baryon number density, is taken to be $Y_L=0.3$~\cite{Raffelt:1996wa}.
The occupancy of $\Lambda$ states is distributed according to the Fermi distribution $f^\Lambda_{\pmb{p}}=1/\big(1+\exp\big({\frac{E_{\Lambda}-\mu_\Lambda}{T}}\big)\big)$
where $\pmb{p}$ is the $\Lambda$ three-momentum in the star's rest frame, $E_\Lambda$ its energy, $E_\Lambda^2=\pmb{p}^2+m_\Lambda^2$, and $\mu_\Lambda$ its chemical potential. Neutrons are distributed following an analogous distribution, $f^n_{\pmb{p^\prime}}$, characterized by $\mu_n$ and labelled by the corresponding neutron three-momentum $\pmb{p^\prime}$, also in the star's frame. Anti-particles follow identical distributions with the replacement $\mu\to-\mu$, so that for the temperatures expected in a PNS the densities of $\bar \Lambda$ and $\bar n$ are negligible.

The volume emission rate $Q$ inside the PNS due to the process $\Lambda\to n a$ is given by,
\beq
\begin{split}
\label{eq:SN}
Q=&\frac{m_n^3\Gamma(\Lambda\to na)}{\pi^2(m_\Lambda^2-m_n^2)} \int^\infty_0 p \,dp \times \\
&\times\int^{p^\prime_{\rm max}}_{p^\prime_{\rm min}}p^\prime dp^\prime \frac{E_\Lambda-E_n}{E_\Lambda E_n}f^\Lambda_{\pmb{p}}(1-f^n_{\pmb{p^\prime}}),
\end{split}
\eeq
where $p^\prime_{\rm max}$ ($p^\prime_{\rm min}$) is the maximal (minimal)  neutron momentum in the $\Lambda\to n a$ decay, if $\Lambda$ has momentum $p=|\pmb{p}|$
(all in the PNS's rest frame)~\footnote{In the star's rest frame $\Lambda$ is moving in the direction of $\hat{\pmb{p}}$. The maximum (minimum) three-momentum of the neutron in the PNS's rest frame is reached when the neutron recoils in the $\Lambda$'s rest frame in the direction (in the direction opposite) to $\hat{\pmb{p}}$.}. Notice that in the non-relativistic limit where $p,~p^\prime \ll m_\Lambda\sim m_n$, and in the limit of no Fermi blocking of the final state neutrons, this formula reduces to a more familiar form,
\begin{align}\label{eq:SNapp}
Q \simeq n_n (m_\Lambda - m_n) \Gamma(\Lambda\to na)~e^{-\frac{m_\Lambda - m_n}{T}},
\end{align}
where $n_n$ is the number density of neutrons. 

Evaluating the distributions (chemical potentials) for benchmark conditions of $T=30$ MeV and nuclear density $\rho=\rho_{\rm nuc}$, using Eq.~(\ref{eq:SN}), we obtain for the energy loss per unit mass $\epsilon=Q/\rho$,
\begin{align}\label{eq:SNresult}
\epsilon=3.6\times10^{38} \, \frac{\rm erg }{\rm s \, g} \GeV^2 \left(\frac{f_1(0)^2}{|F^{V}_{sd}|^2}+\frac{g_1(0)^2}{|F^{A}_{sd}|^2}\right). 
\end{align}
Setting as the maximal limit on $\epsilon$ the energy lost through neutrino emission one second after the collapse of the supernova SN~1987A, $\epsilon\lesssim10^{19} \, {\rm erg}/{\rm s} \, {\rm g}$~\cite{Hirata:1987hu,Raffelt:1996wa}, one obtains bounds on $|F_{sd}^A|$ and $|F_{sd}^V|$ in the range $10^9$ - $10^{10}$ GeV. 

Our estimates are afflicted by significant uncertainties. Nuclear interactions induce important corrections in the calculation of the number densities~\cite{Prakash:1996xs,Vidana:2018bdi} and there are considerable stellar uncertainties stemming from the complex physics at work in the supernova. Note that the energy loss per unit mass obtained using the approximate formula in Eq.~(\ref{eq:SNapp}) is independent of the structural details of the PNS, except for the temperature. At $T=30$ MeV this leads to an emission rate that is $\sim40\%$ larger than in Eq.~(\ref{eq:SNresult}). More than anything, the emission rate suffers from the uncertainty in the temperature of the central region. Variation of this quantity 
from 20 to 40 MeV changes $Q$ by two orders of magnitude.  Finally,
our bound crucially relies on the validity of the standard scenario for the SN explosion as applied to SN~1987A, which was disputed in a recent 
publication~\cite{Bar:2019ifz}. 

\section{Bounds from Meson Mixing}
\label{sec:meson:mix}

\begin{figure*}[t!]
 \vspace{-0.3cm}
 \begin{minipage}{0.26\textwidth}
  \hspace{-3.9cm}
 $a)$\\[7mm]
  \includegraphics[width=3.6cm]{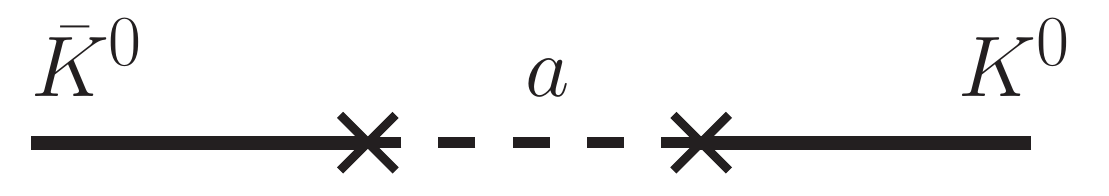} 
\end{minipage}
 \begin{minipage}{0.25\textwidth}
  \hspace{-3.2cm}\vspace{-0.55cm}
 $b)$\\[3mm]
  \includegraphics[width=2.6cm]{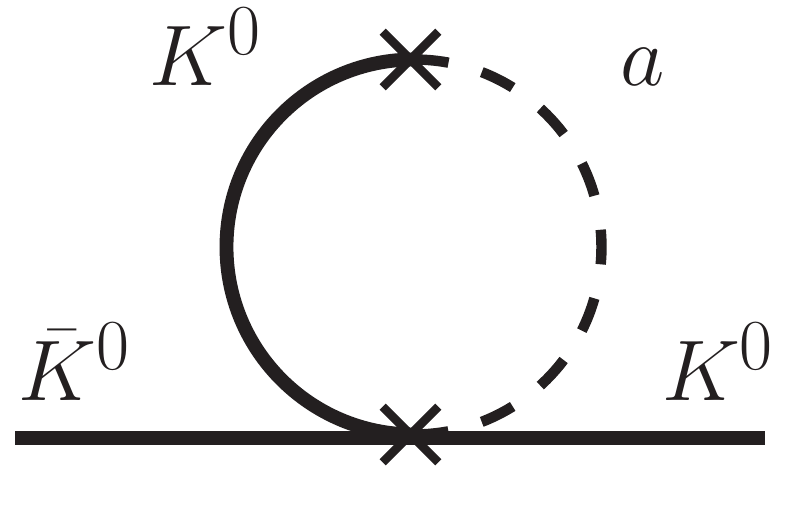} 
\end{minipage}
 \begin{minipage}{0.26\textwidth}
   ${}$\\[8mm]
  \hspace{-3.5cm}\vspace{-.55cm}
 $c)$\\[9mm]
 \vspace{-.15cm}
  \includegraphics[width=3.5cm]{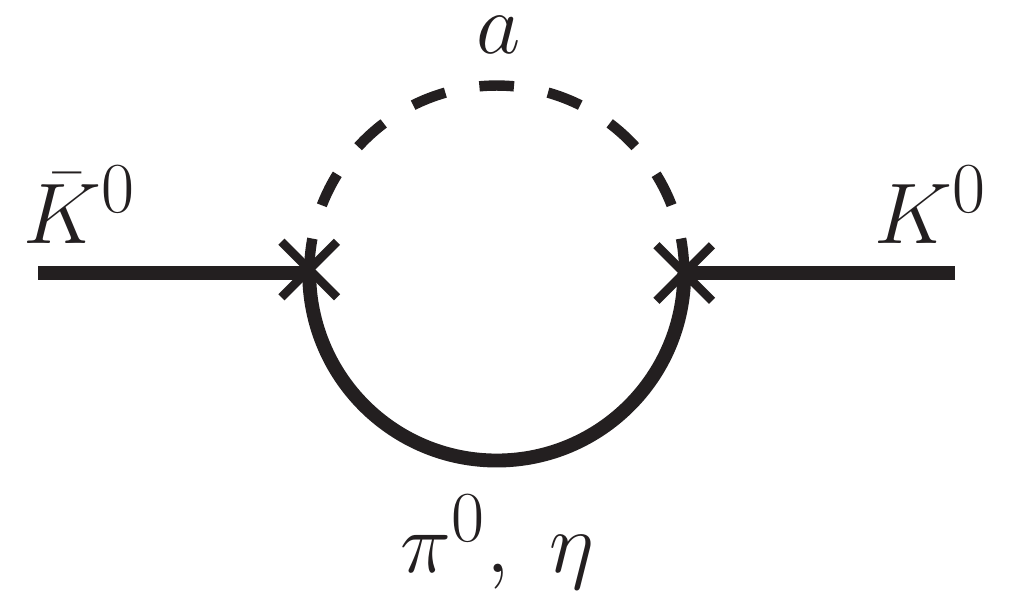} 
\end{minipage}
 \begin{minipage}{0.2\textwidth}
  \hspace{-2.2cm}\vspace{-.55cm}
 $d)$\\[12mm]
  \includegraphics[width=2.1cm]{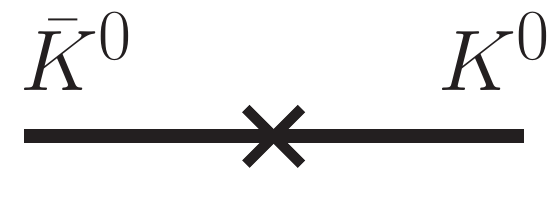} 
\end{minipage}
\caption{Contributions to $K^0$-$\bar K^0$ mixing from exchanges of flavor violating axion, up to and including one loop in ChPT. 
\label{fig:chiral_loops}}
\end{figure*}

The exchanges of axions with flavor violating couplings contribute to $\Delta F=2$ transitions and can modify meson mixing rates from the SM predictions. The contribution from axion exchanges to the mixing amplitude of the $P^0-\bar P^0$ neutral meson system  is given by the time-ordered correlator
\beq\label{eq:def_massdiff}
\begin{split}
M_{12}=- \frac{i}{4 m_P}\int d^4x \langle P^0|T\{\mathcal{L}_{aff} (x),\mathcal{L}_{aff}  (0)\}|\bar P^0\rangle, 
\end{split}
\eeq
where $\mathcal{L}_{aff}(x)$ is the axion-fermion interaction Lagrangian in Eq.~(\ref{eq:couplings}).

In this section it will prove useful to use the following form of the axion-fermion interaction Lagrangian, 
\beq
\begin{split}
\label{eq:couplings:perparts}
 \mathcal{L}_{aff} = -
&i \frac{ a}{2 f_a} \, \bar f_i \big[ (m_{f_i}-m_{f_j})c^V_{f_i f_j} 
\\
&\qquad+ (m_{f_i}+m_{f_j})c^A_{f_i f_j} \gamma_5 \big] f_j  \,, 
\end{split}
\eeq
which is obtained from Eq.~(\ref{eq:couplings}) with the axion-dependent field transformation
\begin{align}
f_i \to \left[ e^{i \frac{a}{2 f_a} (c^V_{ij} - c^A_{ij})}  P_L + e^{i \frac{a}{2 f_a}(c^V_{ij} + c^A_{ij})}  P_R  \right] f_j \, .
\end{align}
Notice that ${\mathcal O}(a^2)$ terms  that also appear in the transformation from Eq.~(\ref{eq:couplings}) to Eq.~(\ref{eq:couplings:perparts}) do not affect $\Delta F=2$ processes, and are omitted for that reason. 
The above form of $\mathcal{L}_{aff}$ simplifies somewhat  the calculations of the non-local meson mixing matrix elements,  Eq.~(\ref{eq:def_massdiff}). For this we utilize the appropriate effective field theories; we use Chiral Perturbation Theory (ChPT) for contributions to $K-\bar K$ mixing, while for the heavy-quark systems, $B_{d,s}-\bar B_{d,s}$  and $D-\bar D$, we use the Operator Product Expansion (OPE), matching onto local four-quark operators describing the meson mixing.  In the following we use the relativistic normalization of states,  $\langle P^0(k)| P^0(k')\rangle=2E\delta(\vec k-\vec k')$, and the phase convention $CP|P^0\rangle=-|\bar P^0\rangle$.

\subsection{$K-\bar K$ mixing}

Since $m_{a}\ll 1{\rm~GeV}$ we can use ChPT to describe contributions from axion exchanges in $K-\bar K$ mixing. For axial couplings, $c^A_{f_if_j}$,
the leading contributions arise at tree level, while for vector couplings, $c^V_{f_if_j}$, the first nonzero contributions are at one loop. 

To construct the ChPT Lagrangian in the presence of flavor violating axions we use a spurion analysis~\cite{Gasser:1984gg,Pich:1995bw}. In terms of the (pseudo)scalar interactions, the Lagrangian for QCD with a flavor violating axion can be conveniently written as
\beq\label{eq:QCDaxionLag}
\begin{split}
{\cal L}_{{\rm QCD}+a}= &\bar q (i \slashed \partial + g_s \slashed G^a T^a)q-\bar q {\cal M}_q q
\\
& -a \,  \bar q ( \chi_S- i \chi_P  \gamma_5) q,
\end{split}
\eeq
where we keep only light quarks, $q=(u,d,s)$. The diagonal mass matrix is ${\cal M}_q=\diag(m_u,m_d,m_s)$,  while $\chi_{S,P}$ are $3\times3$ Hermitian matrices describing the quark-axion couplings,
\begin{align}
\label{eq:chiS}
\chi_{S}&=i \begin{pmatrix}
0 & 0 & 0
\\
0 & 0& \frac{m_d-m_s}{F_{ds}^{V}}
\\
0 &\frac{m_s-m_d}{F_{sd}^{V}} & 0
\end{pmatrix},
\\
\label{eq:chiP}
\chi_{P}&= -\begin{pmatrix}
\frac{2m_u}{ F_{uu}^{A}}& 0 & 0
\\
0 & \frac{2m_d}{ F_{dd}^{A}}& \frac{m_d+m_s} {F_{ds}^{A}}
\\
0 & \frac{m_s+m_d}{ F_{sd}^{A} }& \frac{2m_s}{F_{ss}^{A}}
\end{pmatrix}.
\end{align}
The off-diagonal couplings in \eqref{eq:chiS}, \eqref{eq:chiP} induce kaon oscillations. The mixing matrix element $M_{12}$ follows from a double insertion of the interaction Lagrangian $\mathcal{L}_{aff}=-a \,  \bar q ( \chi_S- i \chi_P  \gamma_5) q$, where $q=u,d,s$, cf. Eq.~(\ref{eq:def_massdiff}).
 
The Lagrangian for QCD  with the flavor violating axion, ${\cal L}_{{\rm QCD}+a}$, is formally invariant under $SU(3)_R\times SU(3)_L$ transformations, $q_{R,L}\to g_{R,L}(x) q_{R,L}$, if $a\chi_{S,P}$ and ${\cal M}_q$ are promoted to spurions that transform as 
\begin{align}
s+ip &\to g_R(s+ip) g_L^\dagger,
\end{align}
and that take the values  
\beq
 s={\cal M}_q+\chi_S a,\, \qquad p=\chi_P a.
\eeq 
We also define the spurion $\chi=\chi_S+i \chi_P$ that does not contain the axion field, and which transforms similarly as $\chi\to g_{R} \chi\,g_{L}^\dagger$. The identification of this symmetry structure allows one to build the ChPT Lagrangian, including the chiral-symmetry breaking terms. The introduction of the spurion $\chi$ is needed because the axion cannot be treated merely as an external field and enters in the chiral loops with two insertions of $\chi_{S,P}$. Thus, the ChPT Lagrangian is also invariant under a $Z_2$ symmetry that transforms $\chi\to -\chi$, with all the other fields taken to be $Z_2$ even.

Up to overall normalization factors, the axion induced contributions to the $K-\bar K$ mixing amplitude have the scalings $ 2 m_K M_{12}\sim (p/\Lambda_\chi)^{\nu} (1/F)^{\nu_F}$. The integer $\nu$ characterizes the usual chiral scaling~\cite{Gasser:1984gg,Pich:1995bw} where the derivatives of meson fields (and the axion) count as $\mathcal O(p)$, the quark masses as $\mathcal O(p^2)$ and where the UV cut-off of ChPT is $\Lambda_\chi\simeq 4\pi f_\pi\sim {\mathcal O}(1{\rm~GeV})$ ($f_\pi$ is the pion decay constant). On the other hand, $\nu_F$ counts the number of $1/F^{V,A}_{ds}$ insertions in the amplitude. Thus, the chiral counting of the spurions is ${\cal M}_q\sim {\mathcal O}(p^2)$ and $\nu_F=0$ and $\chi_{S,P} \sim {\mathcal O}(p^2)$ and $\nu_F=1$.  In the following we use ChPT to calculate the leading axion-exchange contributions to the $K-\bar K$ mixing amplitude
including corrections up to NLO corrections in the chiral counting, $\nu\leq4$. This requires two insertions of $1/F^{V,A}_{ds}$ and thus $\nu_F=2$. 

The LO ChPT Lagrangian, including $a$ as the light degree of freedom, is given by
\beq\label{eq:coupl:LO}
\begin{split}
{\cal L}_{{\rm ChPT}+a}^{(2)} &= \frac{f^2}{4} {\rm Tr}\big(\partial_\mu U \partial^\mu U^\dagger \big)
\\
&+B_0 \frac{f^2}{2} \Tr\big[ (s-ip) U+ (s+ip) U^\dagger\big]
\\
&+\frac{1}{2}\partial_\mu a\partial^\mu a-\frac{m_{a}^2}{2} a^2,
\end{split}
\eeq
while the relevant terms in the NLO ChPT Lagrangian are 
\beq\label{eq:coupl:NLO}
\begin{split}
{\cal L}_{{\rm ChPT}+a}^{(4)} &\supset B_0^2 f^2\Big(\alpha_0 \Tr\big[\chi U^\dagger\big]\Tr\big[\chi^\dagger U\big]
\\
&+\alpha_1 \Tr\big[(\chi U^\dagger)^2+(\chi^\dagger U)^2\big]\Big).
\end{split}
\eeq
Here $U(x)=\exp(i \lambda^a \pi^a/f)$ is the unitary matrix parametrizing the meson fields~\cite{Gasser:1984gg,Pich:1995bw},   
$B_0$  is a constant related to the quark condensate, $B_0(\mu=2{\rm~GeV})=2.666(57)~{\rm GeV}$,  $f$ is related to the pion decay constant $f\simeq f_\pi/\sqrt{2}=92.2(1)$ MeV~\cite{Aoki:2019cca}, with normalization $\langle 0| \bar u \gamma_\mu d(0) |\pi^-(p)\rangle=i p_\mu f_\pi$, and $\alpha_{0,1}\sim f^2/\Lambda_\chi^2$ 
are the (unknown) low energy constants. 

Expanding \eqref{eq:coupl:LO} in meson fields gives 
\beq\label{eq:coupl:LO:expand}
\begin{split}
{\cal L}_{{\rm ChPT}+a}^{(2)} &\supset -\frac{a}{F_{ds}^A} f_K m_K^2 \bar K^0
\\
&- \frac{i}{\sqrt2}  \frac{a}{F_{ds}^V} \big(m_K^2-m_\pi^2\big) \bar K^0\Big(\pi^0+\tfrac{\eta}{\sqrt3} \Big)
\\
&+\frac{a}{F_{ds}^A} \frac{2m_K^2}{3 f_K}  K^0  \big(\bar{K}^0\big)^2 +{\rm h.c.}
\end{split}
\eeq
Here we kept only the terms relevant for $K-\bar K$ mixing, and replaced $\sqrt2 f$ with the kaon decay constant, $f_K=155.6\pm0.4$ MeV~\cite{Tanabashi:2018oca}, thus capturing part of the SU(3) breaking that corrects our results at higher orders in the ChPT expansion~\cite{Gasser:1984gg}. We also traded the products $B_0 m_q$ for the meson masses squared~\cite{GellMann:1968rz,Gasser:1984gg,Pich:1995bw}.
The two terms in \eqref{eq:coupl:NLO}, expanded in meson fields, are
\beq
\begin{split}\label{eq:coupl:NLO:expand}
{\cal L}_{{\rm ChPT}+a}^{(4)} &\supset 2 \biggr[\biggr(\frac{m_K^2}{F_{ds}^A}\biggr)^2\big(\alpha_0+2 \alpha_1\big)
\\
&-\biggr(\frac{m_K^2-m_\pi^2}{F_{ds}^V}\biggr)^2\big(\alpha_0-2 \alpha_1\big)\biggr]\big(\bar{K}^0\big)^2+\cdots,
\end{split}
\eeq
where again we only keep the terms relevant for $K-\bar K$ mixing. 

The axion exchange contributions to the $K-\bar K$ mixing amplitude due to axial vector couplings are proportional to $(1/F^A_{ds})^2$ and are, up to $\mathcal{O}(p^4)$ 
in the chiral counting, given by 
\beq\label{eq:chiral_axial}
 \begin{split}
M_{12}^A=&\biggr(\frac{f_K}{F_{ds}^A}\biggr)^2 \frac{m_K}{2}\biggr\{1-\frac{2 m_K^2}{f_K^2}(\alpha_0+2\alpha_1)
\\
&+\frac{8}{3}\frac{m_K^2}{16\pi^2f_K^2}\Big(1-\log\Big(\frac{m_K^2}{\mu^2}\Big)\Big)\biggr\}.
\end{split}
\eeq
The first term in the parenthesis is due to the tree level axion exchange, Fig.~\ref{fig:chiral_loops} a), and is induced by the first term in the expanded LO ChPT Lagrangian, Eq.~\eqref{eq:coupl:LO:expand}. The NLO correction is due to the loop diagram in Fig.~\ref{fig:chiral_loops} b), induced by the last term in Eq.~\eqref{eq:coupl:LO:expand}. We use dimensional regularization in the $\overline{\text{MS}}$ scheme and fix the renormalization scale to $\mu=m_K$ in order to minimize the size of the chiral logarithm. The counter-terms in Fig.~\ref{fig:chiral_loops} d) which cancel the $\mu$ dependence of the loops  are provided by the two terms in the expanded ${\mathcal O}(p^4)$ Lagrangian, Eq.~\eqref{eq:coupl:NLO:expand}.
While the numerical values of the low energy constants $\alpha_{0,1}$ are not known, we can estimate their sizes by varying $\mu$ in the loop contribution around its nominal value by a factor of $2$, while artificially setting $\alpha_{0,1}$ to zero. This estimates the ${\mathcal O}(p^4)$ contribution in \eqref{eq:chiral_axial} to be $(17\%\pm23\%)$ of the  ${\mathcal O}(p^2)$ one. 

The vector couplings of the axions first contribute at ${\mathcal O}(p^4)$ through the one loop diagram in Fig.~\ref{fig:chiral_loops} c) with the counterterms in Fig.~\ref{fig:chiral_loops} d),  resulting in
\beq
\begin{split}\label{eq:chiral_vector}
M_{12}^V-i \frac{\Gamma_{12}^V}{2}&=\biggr(\frac{f_K}{F_{ds}^V}\biggr)^2\frac{m_K}{2} \biggr(1-\frac{m_\pi^2}{m_K^2}\biggr)^2 \times
\\
&\times \biggr\{\frac{m_K^2 }{32\pi^2 f_K^2}
 \left(I_0(z_\pi)+\tfrac{1}{3}I_0(z_\eta)\right) 
 \\
 &\qquad+\frac{2 m_K^2}{ f_K^2} \big(\alpha_0-2\alpha_1\big)\biggr\}.
\end{split}
\eeq
Above, we have re-arranged the factors of $f_K$ and $m_K$ to make the dependence on $(1/F_{ds}^V)^2$ as well as the structure of the chiral corrections more transparent. The two-point loop function is  $I_0(z)=2-\log(m_K^2/\mu^2)-z\log z-(1-z)\log(z-1-i0^+)$, where the argument is $z_\phi=m_\phi^2/m_K^2$. The $\Gamma_{12}^V$ receives a contribution from the discontinuity, $\Im I_0(z_\pi)$, i.e., from the on-shell part of the diagram Fig.~\ref{fig:chiral_loops} c) with pion and axion running in the loop. 
The choice $\mu=m_K$ again minimizes the log. However, since the low energy constants $\alpha_{0,1}$ are unknown the predicted $M_{12}^V$ is quite uncertain. In the numerical estimates we use $\alpha_{0,1}=0$ and assign 100\% uncertainty to the resulting estimate for $M_{12}^V$. 

As we will see in the next subsection, for heavy quarks both $s$-channel and $t$-channel exchanges of axions lead to contributions that are parametrically of similar size. However, this is not the case for light quarks. The above ChPT analysis implies that the tree level $s$-channel axion exchange (necessarily proportional to $1/(F_{ds}^A)^2$) is leading in the chiral expansion, while the $t-$channel contribution is subleading. 

Note that in addition to the contributions to $K-\bar K$ mixing from axion exchange, there can be other contributions from UV physics which are parametrically of the same order, $M_{12} \propto (1/F^{V,A}_{ds})^2$. In the numerical analysis we set these UV model dependent contributions to zero, keeping in mind that their presence can modify our numerical results. 

Numerically, Eqs.~(\ref{eq:chiral_axial}) and (\ref{eq:chiral_vector}) give for the contribution of the axion to the neutral-kaon mass difference,
\beq\label{eq:KKbaraxion}
\begin{split}
\Delta m_K/m_K&=0.028(6)~\text{GeV}^2\,\Re\big[1/(F_{ds}^A)^2\big]
\\
&+0.0018(18)~\text{GeV}^2\,\Re\big[1/(F_{ds}^V)^2\big],
\end{split}
\eeq
where we use the relation $\Delta m_K=2{\rm Re} M_{12}$ and have set $\alpha_0=\alpha_1=0$. The prefactor of the vector couplings carries an ${\mathcal O}(100\%)$ relative uncertainty because of the unknown contributions of $\alpha_{0,1}$ coefficients, entering at the same order in ChPT as the loop contribution.
The prediction of $\Delta m_K$ in the SM has large uncertainties stemming from long-distance contributions. 
Therefore, in order to obtain the bounds on the axion couplings from this observable we conservatively saturate the experimental value $\Delta m_K^{\rm expt.}=3.484(6)\times10^{-12}$ MeV~\cite{Tanabashi:2018oca} with the axion-exchange contribution. Assuming $\alpha_{0,1}=0$ this leads to $|F_{ds}^A|>2.0 \cdot 10^{6}$ GeV and $|F_{ds}^V|>5.1 \cdot 10^{5}$ GeV at $90\% $ C.L., for the case where $1/(F_{ds}^{A,V})^2$ are real. These are the bounds quoted in Tab. \ref{tab:main:results}. 
Taking into account the estimate for the range of values for $\alpha_{0,1}$ reduces the bound on $|F_{ds}^A|$ by about 10\%. Note that without fixing the values of $\alpha_{0,1}$ there is no bound on $|F_{ds}^V|$. Allowing for large cancellations up to 1\% between the loop diagram and the counterterm contributions relaxes the bound on $|F_{ds}^V|$ by an order of magnitude. 

To obtain the bounds on non-SM CP violating contributions to $K-\bar K$ mixing we use the normalized quantity
\beq
C_{\varepsilon_K} = \frac{|\epsilon_K^{{\rm SM}+a}|}{|\epsilon_K^{\rm SM}|}.
\eeq
 For the theoretical prediction of $\epsilon_K$ we use the expression \cite{Buras:2010pza}
\beq
\epsilon_K=e^{i\phi_\epsilon}\sin\phi_\epsilon\biggr(\frac{\Im M_{12}}{\Delta m_K}+\xi\biggr),
\eeq
where
\beq
\xi\simeq\frac{\Im \Gamma_{12}}{\Delta \Gamma_K}.
\eeq
We take the values for $\Delta m_K=m_L-m_S$, $\Delta \Gamma_K=\Gamma_S-\Gamma_L$, and $\phi_\epsilon=\arctan(2\Delta m_K/\Delta \Gamma_K)$ from experiment~\cite{Tanabashi:2018oca}. With the SM prediction for $|\epsilon_K|$ from \cite{Brod:2019rzc}, and the axion contributions to $M_{12}$, $\Gamma_{12}$ from Eqs.~\eqref{eq:chiral_axial}, \eqref{eq:chiral_vector} we get 
\beq\label{eq:axion:CeK}
\begin{split}
\delta C_{\epsilon_K}=C_{\epsilon_K}-1&=\Im\biggr[\biggr(\frac{2.5(2)\cdot 10^7{\rm~GeV}}{F_{ds}^A}\biggr)^2
\\
&+\biggr(\frac{5(5)\cdot 10^5{\rm~GeV}}{F_{ds}^V}\biggr)^2\biggr],
\end{split}
\eeq
where in the numerical expressions we set the unknown low energy constants to zero, $\alpha_{0,1}\to 0$, with the quoted errors our estimates of the resulting errors due to this approximation~\footnote{There is a 1:100 cancellation between the contributions of ${\rm Im}M_{12}^V$ and ${\rm Im}\Gamma^V_{12}$ to $\epsilon_K$ in Eq.~\eqref{eq:axion:CeK} that makes the prediction of the vectorial axion couplings to this observable even more uncertain. For the experimental inputs we use the PDG values~\cite{Tanabashi:2018oca}, using $m_{K_0}$ and $m_{\pi_0}$ in \eqref{eq:chiral_vector} for the kaon and pion masses, respectively.}. 

The global CKM fit by the UTFit collaboration obtains $0.87< C_{\epsilon_K}<1.39$  at 95\% CL \cite{Bona:2007vi,UTfitweb}. Assuming $\alpha_{0,1}=0$ this translates into the  90\% CL bounds $|F_{ds}^A|>4.4 (7.7) \cdot 10^7$ GeV, $|F_{ds}^V|>0.9 (1.5) \cdot 10^6$ GeV for purely imaginary and positive (negative) $(1/F_{ds}^{A,V})^2$ which maximally increase (reduce) $C_{\epsilon_K}$ above (below) $1$. 

In Tab. \ref{tab:main:results} we quote the bounds from the less stringent case of CP violating contributions that positively interfere with the SM contributions to $\epsilon_K$.
These bounds will improve in the future, once the improved prediction for $\epsilon_K$~\cite{Brod:2019rzc} is implemented in the global CKM fits.

\subsection{Heavy-meson mixing}

In the mixing of neutral heavy mesons, $B_{s,d}-\bar B_{s,d}$ or $D-\bar D$, a large momentum, of the order of the heavy quark mass, $p\sim m_Q\gg \Lambda_{\rm QCD}$, is injected through ${\cal L}_{aff}$, Eq.~\eqref{eq:def_massdiff}, to an intermediate set of states of the form $a+hadrons$. This allows one to perform an OPE in $x\sim 1/m_Q$ and express the bi-local operator in terms of the local ones,  
\beq
\begin{split}
\label{eq:matching:OPE}
\frac{i}{2}  \int d^4x&T\{{\mathcal L}_{aff}(x),{\mathcal L}_{aff} (0)\}\to
\\
&\to {\cal L}_{\rm eff}(0)=\sum_{i}C_{i}{\mathcal O}^{\Delta F=2}_i(0).
\end{split}
\eeq
The sum runs over different dimensions and possible structures of the local operators. The mass matrix element leading to the meson mixing is 
\beq
M_{12}=- \frac{1}{2 m_P} \langle P^0|\mathcal{L}_{\rm eff} (0)|\bar P^0\rangle. 
\eeq

At LO in the $1/m_Q$ expansion the local operators ${\mathcal O}^{\Delta F=2}_i(0)$ in \eqref{eq:matching:OPE} are of dimension-six.
One may have therefore naively expected the Wilson coefficients to scale as $C_{i}\propto 1/m_Q^2$. However, the axion couplings to quarks are  $\propto m_Q/f_a$, cf. Eq.~\eqref{eq:couplings:perparts},
leading to $C_{i}\propto1/f_a^2$. Note also that we are calculating the OPE at one single kinematic point over the physical cut of the $a+hadrons$ intermediate state. We assume the mass of the heavy-quark to be large enough so that the energies involved correspond to the perturbative regime of QCD and that the violations of quark-hadron duality are small.

 We first present the results for the $B^0-\bar B^0$ system, and then extend the results to $B_s^0-\bar B_s^0$ and $D^0-\bar D^0$ systems. We work at LO in $1/m_Q$. The full basis of local operators at this order is given by~\cite{Ciuchini:1998ix},
\beq
\begin{split}
\label{eq:local_ops}
\mathcal O_1&=(\bar d^\alpha_L\gamma^\mu b_L^\alpha)(\bar d_L^\beta\gamma^\mu b_L^\beta),
\\
\mathcal O_2&=(\bar d^\alpha_R b_L^\alpha)(\bar d^\beta_R b_L^\beta),~~~\mathcal O_3=(\bar d^\alpha_R b_L^\beta)(\bar d^\beta_R b_L^\alpha),\\
\mathcal O_4&=(\bar d^\alpha_R b_L^\alpha)(\bar d^\beta_L b_R^\beta),~~~\mathcal O_5=(\bar d^\alpha_R b_L^\beta)(\bar d^\beta_L b_R^\alpha),
\end{split}
\eeq
along with the operators $\tilde{\mathcal O}_{1,2,3}$ obtained by replacing $P_L\to P_R$  in $\mathcal O_{1,2,3}$.  The summation over color indices,  $\alpha$, $\beta$, is implied. The operator basis for $B_s-\bar B_s$ mixing is obtained from the above by replacing $d\to s$, and for $D-\bar D$ mixing by replacing $b\to c$, $d\to u$.

\begin{figure}[t]
 \includegraphics[width=9.cm]{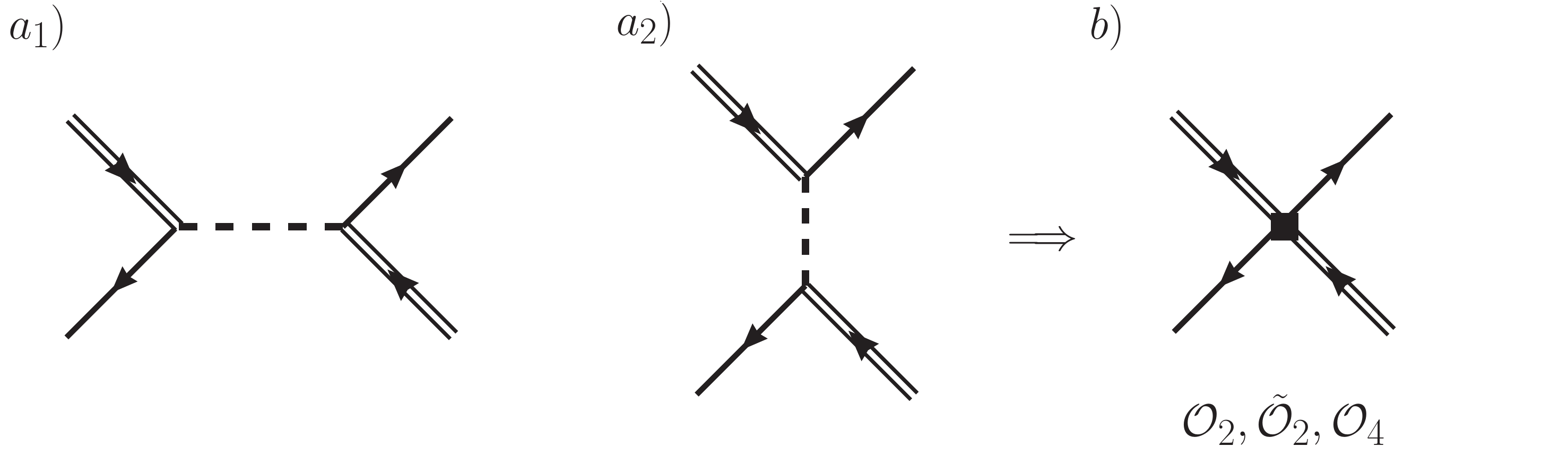} 
\caption{Matching of the axion-mediated contributions to $B-\bar B$ mixing, diagrams $a_1)$ and $a_2)$, onto the dimension-six local operators in the OPE, diagram $b)$, where we also indicate the operators that receive the contributions. Single (double) lines represent the $d$ quark ($b$ quark). 
\label{fig:OPE}}
\end{figure}

The Wilson coefficients $C_i$  are most easily obtained by matching both sides of Eq.~\eqref{eq:matching:OPE} for axion-mediated $b{\bar d}\to \bar b d$ scattering with on-shell quarks in the initial and final state, see Fig.~\ref{fig:OPE}. The axion can be exchanged in $s$- and $t$-channels. In both cases the axion is far off-shell, $p_a^2\sim \mathcal O(m_b^2)$, where $m_b\gg \Lambda_{\rm QCD}$, justifying the application of the OPE. Note that for the virtuality of the axion in the $t$-channel we are using the fact that in the heavy-quark limit $p_{B^0}=p_{\bar b}$ and $p_{\bar B^0}=p_b$.

The matching at ${\mathcal O}(\alpha_S^0)$ leads to the following nonzero Wilson coefficients, 
\begin{align}
\label{eq:WCs_OPE:C2}
C_2&=\frac{1}{2}\left(\frac{1}{F^A_{db}}+\frac{1}{F^V_{db}}\right)^2,
\\
\label{eq:WCs_OPE:tildeC2}
\tilde C_2&=\frac{1}{2}\left(\frac{1}{F^A_{db}}-\frac{1}{F^V_{db}}\right)^2,
\\
\label{eq:WCs_OPE:C4}
C_4&=\frac{1}{(F^V_{db})^2}-\frac{1}{(F^A_{db})^2},
\end{align}
and similarly for the $B_s$ system, with $d\to s$. 
The matching gets corrected at $\mathcal O (\alpha_s)$ due to hard-gluon contributions to the matching, and at $\mathcal O(\Lambda_{\rm QCD}/m_b)$ from corrections to the heavy-quark limit.

The matrix elements of the operators in Eq.~(\ref{eq:local_ops}) between $B^0$ and $\bar B^0$ states are given by $\langle B^0|\mathcal O_i|\bar B^0\rangle\propto f_B^2 B_i$, where  $f_B$ is the $B$ meson decay constant, and $B_i$ the appropriate bag parameter. Both of these are well known and have been
 been calculated using lattice QCD~\cite{Bazavov:2012zs,Carrasco:2013zta,Bazavov:2016nty,Dowdall:2019bea}. 
Parity conservation of strong interactions implies $\langle \tilde O_i \rangle=\langle O_i \rangle$ so that the contributions to $\Delta m_B$ from the $\pm1/(F_{db}^AF_{db}^V)$ terms in \eqref{eq:WCs_OPE:C2} and \eqref{eq:WCs_OPE:tildeC2} cancel. Using the lattice results from Ref.~\cite{Dowdall:2019bea} we find
\begin{align}\label{eq:massB}
\frac{ \Delta m_B}{m_B}&={\rm Abs}\left[\frac{0.053(5)~\text{GeV}^2}{(F_{db}^A)^2}-\frac{0.016(2)~\text{GeV}^2}{(F_{db}^V)^2}\right],
\nonumber \\
\frac{\Delta m_{B_s}}{m_{B_s}}&={\rm Abs}\left[\frac{0.077(8)~\text{GeV}^2}{(F_{sb}^A)^2}-\frac{0.020(2)~\text{GeV}^2}{(F_{sb}^V)^2}\right],
\end{align}
where $\Delta m_{B_{(s)}}=2|M_{12}^{(bs)}|$ and with the dominant theoretical uncertainty due to power-corrections entering at $\Lambda_{\rm QCD}/m_b\sim0.1$. The reduced sensitivity to the vectorial couplings in this formula was anticipated already in the vacuum-insertion approximation estimate~\cite{Feng:1997tn}. The SM predictions for $\Delta m_{B_{d,s}}$ are consistent, within $\sim 10\%$, with the measured values $\Delta m_{B_d}=3.354(22)\times10^{-10}$ MeV and $\Delta m_{B_s}=1.1688(14)\times10^{-8}$ MeV~\cite{Tanabashi:2018oca}. Comparison with our predictions immediately shows that we can expect the bounds on $F_{sb,db}^{V,A}$ at the level of $10^5$ to $10^6$ GeV.

To derive the precise bounds on the allowed axionic contributions we consider simultaneously both the contributions to $\Delta m_{B_{d,s}}$ as well as to the mixing phase. We thus define
\beq
C_{B_q} e^{2 i\phi_{B_q}}=\frac{\langle B_q^0|{\cal L}_{\rm eff}^{{\rm SM}+a}|\bar B_q^0\rangle}{\langle B_q^0|{\cal L}_{\rm eff}^{\rm SM}|\bar B_q^0\rangle}, \quad (q=d,s).
\eeq
In the SM $C_{B_q}=1$ and $\phi_{B_q}=0$. From the global fit to CKM observables, including $B_q-\bar B_q$ mixing observables, the UTFit collaboration obtained $C_{B_d}=1.05\pm0.11$, $\phi_{B_d}=(-2.0\pm1.8)^\circ$, $C_{B_s}=1.110\pm0.090$, $\phi_{B_s}=(0.60\pm0.88)^\circ$ ~\cite{Bona:2007vi,UTfitweb}. The SM predictions are obtained using the inputs in Table \ref{tab:inputs} and ref.~\cite{Dowdall:2019bea}, and the results for the CKM matrix elements of the ``New-Physics fit'' of the UTfit collaboration (Summer 2018)~\cite{Bona:2007vi,UTfitweb}. This leads to the 90\% C.L. bounds
  \begin{align}
 | F_{db}^V|&>1.1(1.3)\cdot 10^6 {\rm~GeV},\\
 | F_{sb}^V|&>{2.0 (3.8)}\cdot 10^5{\rm~GeV},
  \end{align}
  and
    \begin{align}
  |F_{db}^A|&>2.0(2.3)\cdot 10^6 {\rm~GeV},\\
  |F_{sb}^A|&>4.0(7.7)\cdot 10^5{\rm~GeV},
  \end{align}
  when the weak phase of $F_{qb}^{A,V}$ is aligned with (differs by $\pm \pi/2$ from) the SM contribution. The first choice corresponds to MFV like couplings of the axion, where the CPV phase is the SM one, while the second choice corresponds to generic couplings with new weak phase that maximizes the axion exchange contribution to the meson mixing phase. In deriving the bounds above we chose in each case the sign of the NP contribution that leads to the weakest bound. For each of the bounds we also assume that the axion only has axial or vector couplings. These bounds are also collected in Tab.~\ref{tab:main:results}.

Extending the above OPE results to the $D^0-\bar D^0$ system could be problematic since the violations of quark-hadron duality in the $a+hadrons$ system at $\sqrt{s}\simeq m_{D^0}$  may be large. Extending naively our results we obtain 
for the axion contribution to the mass difference
\beq
\begin{split}
\label{eq:DeltamD}
\frac{\Delta m_D}{m_D}&=\frac{2|M_{12}^{cu}|}{m_D}=\frac{1}{m_D^2}|\langle D^0|\mathcal{L}_{\rm eff}^a (0)|\bar D^0\rangle|
\\
&={\rm Abs}\left[\frac{0.12~\text{GeV}^2}{(F_{uc}^A)^2}-\frac{0.034~\text{GeV}^2}{(F_{uc}^V)^2}\right].
\end{split}
\eeq
In the last line we used the lattice QCD results for the bag parameters from~\cite{Bazavov:2017weg} (see also~\cite{Carrasco:2014uya,Carrasco:2015pra}). To obtain a bound on $|F^{A,V}_{uc}|$ we saturate the experimental value of $\Delta m_D$ with the axion contribution because the SM prediction is poorly known. 

A more stringent bound is obtained for CP violating axionic contributions  from an experimental bound on the $D-\bar D$ mixing phase \cite{Kagan:2020vri}
\beq
\phi_{12}\equiv  \phi_2^M - \phi_2^\Gamma,
\eeq
where in the most commonly used phase convention 
\beq
\phi_2^M={\rm arg}(M_{12}), \qquad  \phi_2^\Gamma={\rm arg}(\Gamma_{12}).
\eeq
The axion 
contributes at tree level to $M_{12}$ and only at loop level to $\Gamma_{12}$. For present experimental bounds the SM contributions to $\phi_{12}$ are negligible, i.e., $\phi_{12}$ at present experimental levels would be induced entirely by the axion exchanges, and thus
\beq
\label{eq:phi12}
\phi_{12}\simeq \phi_2^M =\frac{\Im M_{12}}{|M_{12}|}\simeq \frac{2 \Im M_{12}^a}{\Delta m_D}=\frac{2 \Im M_{12}^a}{x\,\Gamma}.
\eeq
Above we shortened $M_{12}^{cu}\to M_{12}$, while $M_{12}^a$ is the mixing matrix element due to the axion exchange. 

Comparison with the experiment, the HFLAV Moriond 2019 average $\phi_{12}=-(0.25\pm0.97)^\circ$ \cite{Amhis:2019ckw}, and PDG value for mass difference $\Delta m_D=(95\pm43)\cdot 10^8 \hbar s^{-1}$, gives at 90\% C.L.,
 \begin{align}
  |F_{cu}^V|&>0.2(2.5)\cdot 10^7 {\rm~GeV},\\
  |F_{cu}^A|&>0.5(4.8)\cdot 10^7{\rm~GeV},
 \end{align}
 when saturating $\Delta m_D$ ($\phi_{12}$) with the axion contribution, Eq.~\eqref{eq:DeltamD} (Eq.~\eqref{eq:phi12}) and where in Eq.~\eqref{eq:phi12} we used the central value of $\Delta m_D$. 
 
 At the end of LHCb Upgrade II the experimental constraints are expected to reach the parametric sizes of the SM contributions to the two mixing phases, $\phi_2^M\sim \phi_2^\Gamma\sim 10^{-3}$. For the projections of future sensitivities we thus still use the projected 90\% CL bound on $|\phi_2^M|<2.0 \cdot 10^{-3}$ (approximate universality fit projection in \cite{Kagan:2020vri}), assuming that the axion saturates the upper bound, i.e., we assume no cancellations with the poorly known SM contribution. This gives for the expected future sensitivities $ |F_{cu}^V|>7.8 \cdot 10^7 {\rm~GeV}$, $|F_{cu}^A|>1.4\cdot 10^8{\rm~GeV}$, while the bounds from $\Delta m_D$ do not change.

Finally, we reiterate that the above mixing bounds on $F^{V,A}_{qq'}$ assume that the axion contribution is not cancelled by the UV contributions from heavy particles present in the UV theory, even though the latter are expected to be parametrically of the same size.


\section{Bounds on Flavor-conserving Axion Couplings} 
\label{sec:diag}
For completeness we briefly review the constraints on flavor-conserving axion couplings that are dominated by astrophysical bounds from star cooling. These constrain axion couplings to photons, electrons and nucleons, defined by the Lagrangian 
\begin{align}
{\cal L} & = \frac{\alpha_{\rm em}}{8 \pi} \frac{a}{f_a} C_\gamma F^{\mu \nu} \tilde{F}^{\mu \nu} +  \frac{\partial^\mu a}{2 f_a} C_e \, \overline{e} \gamma_\mu \gamma_5 e \nonumber \\
& +   \frac{\partial^\mu a}{2 f_a} C_p \, \overline{p} \gamma_\mu \gamma_5 p +  \frac{\partial^\mu a}{2 f_a} C_n \, \overline{n} \gamma_\mu \gamma_5 n \, .
\end{align}
For a wide range of axion masses, the strongest bound on axion coupling to photons $F_\gamma \equiv f_a/C_\gamma \geq 1.8 \times 10^{7} \GeV$ (95\% CL) is set both by the CAST experiment~\cite{Anastassopoulos:2017ftl} and the  evolution of Horizontal Branch (HB) stars in globular clusters~\cite{Ayala:2014pea}. The CAST successor, IAXO, is expected to improve this bound by about an order of magnitude~\cite{Irastorza:2018dyq, Armengaud:2019uso}. For restricted ranges of DM axion masses close to $m_a\sim 3 \, \mu$eV the ADMX experiment probes the axion-photon couplings up to $F_\gamma \gtrsim 10^{12} \GeV$~\cite{Braine:2019fqb}. Note that the photon coupling $C_\gamma = E/N - 1.92(4)$ can be suppressed only by tuning  the ratio of EM and color anomaly coefficients, and thus is expected to be ${\cal O}(1)$ in the bulk of UV axion models.

Axion couplings to electrons are constrained from star cooling, specifically from their impact on the luminosity function of White Dwarfs (WD) as assessed in Ref.~\cite{Bertolami:2014wua}, giving $F_e \equiv 2 f_a/C_e \ge 4.9 \, (4.6) \times 10^{9} \GeV$ at 90 (95)\% CL.  Interestingly, there are hints of anomalous energy loss in stars that may be explained by axions with non-zero couplings to electrons (and possibly photons) of the order of $F_e\approx 7 \times 10^9 \GeV$~\cite{Giannotti:2015kwo, Giannotti:2017hny}.  

Finally, axion couplings to nucleons are constrained by axion emission from the core of supernovae. Converting the results of  Ref.~\cite{Carenza:2019pxu} to our notation gives the bound $F_N \equiv 2 f_a/C_N \gtrsim 1.0 \times 10^{9} \GeV$, where the effective coupling to nucleons, $C_N$, is given in terms of axion couplings  to protons and neutrons as $C_N^2 = {C_n^2 + 0.29 \, C_p^2 + 0.27 \, C_p C_n}$. The nucleon couplings are related to the quark couplings (taking a UV scale of $10^{12} \GeV$ and including only QCD running effects)~\cite{Villadoro, DiLuzio:2017ogq}
\begin{align} 
C_p +C_n &= 
0.50(5)\left( c^A_{uu}  + c^A_{dd}  -1\right) - 2 \delta  \, ,  \\
C_p -C_n &= 1.273(2)\left( c^A_{uu}  - c^A_{dd}  - {\frac{1-z}{1+z}} \right) \, ,
\label{CpmCn}
\end{align}
where $z = m_u/m_d = 0.48(3)$ and $\delta \equiv 0.038(5) c^A_{ss} + 0.012(5) c^A_{cc} +  0.009(2)  c^A_{bb} + 0.0035(4) c^A_{tt}$. Similar to axion-photon couplings, the couplings to protons and neutrons can be suppressed only by tuning, see Refs.~\cite{DiLuzio:2017ogq, Bjorkeroth:2019jtx, Saikawa:2019lng}. However, as discussed in Section~\ref{sec:bounds:SN} this bound relies on the validity of the standard scenario for the SN explosion.

\section{Axion couplings to top quarks}
\label{sec:top:quark}

\begin{figure}[t!]
  \includegraphics[width=6cm]{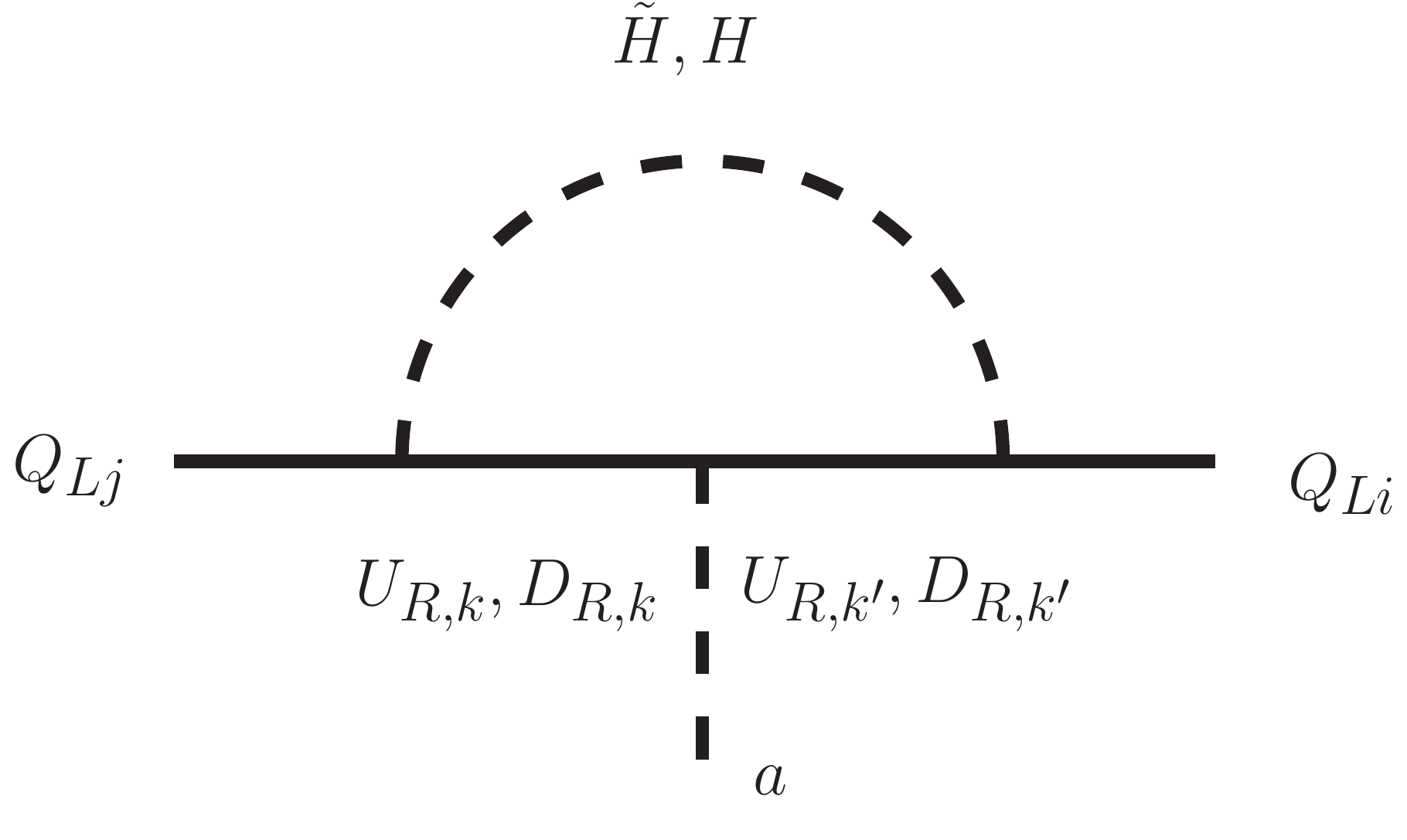} 
\caption{Radiative contribution to the flavor mixing of the couplings of the axion to the quark-doublets.    
\label{fig:loop_run}}
\end{figure}

Finally we discuss flavor-violating couplings of the axion that involve the top quark. Direct bounds on flavor-violating axion-top couplings are, in principle, accessible at the LHC. Monotop searches~\cite{Andrea:2011ws,Boucheneb:2014wza} are sensitive to the $t\to ca, ua$ FCNC transitions, and $t\bar t$+MET~\cite{Arina:2016cqj} to the diagonal $\bar t t a$ couplings. 
However, these searches  bound $f_a$ only very weakly, at the level  
of the electroweak scale. Namely, simply restricting the contribution of $t\to ca$ to the total width of the top to be smaller than ${\mathcal O}(1\text{ GeV})$, gives a bound $f_a\gtrsim {\mathcal O}(v_{\rm EW})$ while the mono-top searches~\cite{Boucheneb:2014wza} may lead to a bound on $f_a$ that is roughly an order of magnitude stronger. 

Much more stringent bounds on top couplings to axions can be obtained from virtual corrections. Because of the large top mass, the radiative yukawa corrections from axion-top couplings (Fig.~\ref{fig:loop_run}) can give sizable contributions to other axion-fermion couplings that are strongly constrained. The leading log expressions for the radiative contributions are derived in Appendix~\ref{sec:app:RGEs}, with the $y_t$-enhanced contributions collected in Eqs.~\eqref{DeltaytA}-\eqref{DeltaytB}. 

The most relevant effects are the radiative corrections to the flavor-violating coupling $c^{V}_{sd}$, subject to  stringent constraints from $K \to \pi a$ (cf. Section~\ref{sec:bounds:2bodymeson:decays}), and the flavor conserving coupling to electrons, $c^A_{ee}$, which is constrained by WD cooling (cf. Section~\ref{sec:diag}). The $y_t^2$ enhanced contributions to these couplings are 
\begin{align}
\Delta c^{V}_{sd} (\mu) & =   \frac{y_t^2}{64 \pi^2} \log\frac{f_a}{\mu}  \left[ 2 V^*_{ts} V_{td} (c^V_{ tt} + c^A_{tt} )  \right. \nonumber \\
& \left. - \sum_k  V^*_{ks} V_{td}   (c^V_{u_k t} - c^A_{u_k t} )   \right.  \nonumber \\
& \left. - \sum_k V^*_{ts} V_{kd} (c^V_{t u_k} - c^A_{t u_k} )   \right] \, , 
\label{eq:DeltacVsd}\\
\Delta c^A_{e e} (\mu) & =   \frac{6 y_t^2}{16 \pi^2} \log\frac{f_a}{\mu}   c^A_{tt}  \, ,
\end{align}
where $\mu$ is the low-energy scale, while on the right-hand side the couplings are given at the UV scale $f_a$.

These contributions are added to the tree-level  $sd-a$ and $ee-a$ couplings, so that from observations we can bound only the sum of the loop induced and tree level coupling, $c_{ij}(\mu)=\Delta c_{ij}(\mu)+c_{ij}(f_a)$. Barring cancellations between the two contributions one can thus obtain bounds on the top quark couplings to axions. The UV values of $sd-a$ and/or $ee-a$ couplings can  be suppressed in certain models. 
A suppressed $sd-a$ coupling arises in scenarios where the down-quark sector is aligned, or all flavor violating axion couplings are strongly suppressed~\cite{Frere:1981cc,Freytsis:2009ct,Dolan:2014ska,Batell:2009jf,Gavela:2019wzg}. A scenario with suppressed $ee$ couplings instead arises in, e.g., DFSZ-type models where all charged lepton couplings are suppressed if the ratio of the Higgs vevs is small. 

Assuming that indeed $\Delta c_{ij}(\mu)\gg c_{ij}(f_a)$, the radiative contributions to FCNCs ($\Delta c^V_{sd}$) and WD cooling ($\Delta c^A_{ee}$) give stringent bounds on top-axion couplings.

The strongest constraint on the $sd$ coupling, $F^V_{sd} \gtrsim 6.8 \times 10^{11}$ GeV translates to a bound on the diagonal top-axion coupling $F^{A}_{tt}\gtrsim 2.5 \times  10^7$ GeV, as well as the off-diagonal couplings $F^{V,A}_{tc}\gtrsim 3.2 \times 10^8$ GeV and $F^{V,A}_{tu}\gtrsim 7.0 \times 10^8$ GeV. Here we have assumed real couplings for simplicity (purely imaginary couplings result in a slightly different bound), and set in \eqref{eq:DeltacVsd} $y_t=y_t ^{\rm SM}(\mu = M_Z)$, $f_a = 10^{10} \GeV$ and used the values of the CKM elements of the ``New-Physics fit'' of the UTfit collaboration (Summer 2018)~\cite{Bona:2007vi,UTfitweb}. With the same numerical inputs one can derive a bound on the diagonal top couplings  from WD cooling~\cite{Feng:1997tn} using the bound $F_{e} \ge 4.9 \times 10^{9} \GeV$ at 90\% CL. This  translates to $F^{A}_{tt}\gtrsim 3.4 \times 10^9$ GeV, which is about two orders of magnitudes stronger than the bound from $K \to \pi a$. Note that similar radiative contributions are obtained for diagonal light quark couplings which can have an impact in the bounds derived from supernovae in specific models.

\section{Results} 
\label{sec:results}

\begin{table}[t]
\begin{tabular}{ccccc}
\hline\hline
Flavors &Process&$F_{ij}^V$ [GeV]&$F_{ij}^A$ [GeV]&Ref.\\
\hline\\[-3mm]
\multirow{16}{*}{$s\to d$}&$K^+\to\pi^+a$& ${\bf 6.8\times 10^{11}}$& --&\cite{Adler:2008zza}\\
			  &&${\bf (2 \times 10^{12})}$&--&\\
                         &$K^+\to\pi^+\pi^0a$ & --&$1.7\times 10^{7}$&\cite{E391a:2011aa}\\
                         &&&(${\bf  7 \times10^8}$)&\\
                         &$\Lambda\to n~a~\text{(decay)}$&$6.9\times 10^{6}$ &$5.0\times 10^6$&\cite{Tanabashi:2018oca}\\
                         &&$(1 \times 10^9)$& ${\bf (8 \times 10^8)}$\\
                         &$\Lambda\to n~a~\text{(SN)}$&$7.4\times 10^9~^\dagger$ &${\bf 5.4\times 10^9~^\dagger}$&\\ 
                         &$\Sigma^+\to p a$&$6.7\times 10^6$&$2.3\times 10^6$&\cite{Tanabashi:2018oca}\\
                         &                 &$(7 \times 10^8)$&$(3 \times 10^8)$&\\
                         &$\Xi^-\to\Sigma^- a$&$1.0\times 10^7$&$1.3\times 10^7$&\cite{Tanabashi:2018oca}\\ 
                         &$\Xi^0\to\Sigma^0 a$&$1.6\times 10^7$&${\bf 2.0\times 10^7}$&\cite{Tanabashi:2018oca}\\ 
                         &                 &$(2 \times 10^8)$&$(3 \times 10^8)$&\\
                         &$\Xi^0\to\Lambda a$&$5.4\times 10^7$&$1.0\times 10^7$&\cite{Tanabashi:2018oca}\\ 
                         &                 &$(9 \times 10^8)$&$(2 \times 10^8)$&\\
                         &$K-\bar K$ ($\Delta m_K$)&$5.1\times 10^5~^\dagger$&$2.0\times 10^6$&\cite{Tanabashi:2018oca}\\
                         &~~~~~~~~~~~~($\epsilon_K$)&$0.9\times 10^6~^\dagger$&${\bf 4.4\times 10^7}$&\cite{UTfitweb}\\
\hline\\[-3mm]
\multirow{7}{*}{$c\to u$}&$D^+\to \pi^+ a$&${\bf 9.7\times 10^7}$&--&\cite{CLEO2008}\\
			&&${\bf (5 \times 10^8)}$&--\\
                         & $\Lambda_c\to p~a$&$1.4\times 10^5$&$1.2\times 10^5$&\cite{Tanabashi:2018oca}\\
                         &&$(2 \times 10^7)$&${\bf (2 \times 10^7)}$\\
                         & $D-\bar D$ (CP cons.)& $2.4\times 10^6~^\dagger$& $4.6\times 10^6~^\dagger$&\cite{Tanabashi:2018oca}\\
                         & ~~~~~~~~~~(CP viol.)& $2.5\times 10^7$& ${\bf  4.8\times 10^7}$& \cite{Amhis:2019ckw}\\         
                         & ~~~~~~~~~~~~~~~~~& $(8 \times 10^7)$& ${\bf (1 \times 10^8)}$& \cite{Kagan:2020vri}\\                  
\hline\\[-3mm]
\multirow{9}{*}{$b\to s$}&$B^{+,0}\to K^{+,0} a$&${\bf 3.3\times 10^8}$&--&\cite{BABAR13}\\
                          &&(${\bf 3 \times 10^{9}}$)&--\\
                          &$B^{+,0}\to K^{*+,0} a$&--&${\bf 1.3\times 10^8}$&\cite{BABAR13}\\
                          &&--&${\bf (1 \times 10^{9})}$\\
                         & $\Lambda_b\to \Lambda~a$&$2.1\times 10^6$&$1.4\times 10^6$&\cite{Tanabashi:2018oca}\\
                         & $B_s \to \mu^+  \mu^-  a$&--&$2.2\times 10^5$&\cite{Albrecht:2019zul}\\
                         & &&$(9 \times 10^5$)\\
                         & $B_s-\bar B_s$ (MFV)& $2.0\times 10^5$ & $4.0\times 10^5$&\multirow{2}{*}{\cite{UTfitweb}}\\
                         & ~~~~~~~~~~~(gen.) &$3.8\times 10^5$ & $7.7\times 10^5$&\\                         
\hline\\[-3mm]
\multirow{8}{*}{$b\to d$}&$B^{+}\to \pi^{+} a$&${\bf 1.1\times 10^8}$& --&\cite{Aubert:2004ws}\\
			& &${\bf (3 \times 10^9)}$& --\\
                         &$B^{+,0}\to \rho^{+,0} a$&--&(${\bf 1 \times 10^9}$)\\
                         & $\Lambda_b\to n~a$&$3.1\times 10^6$&${\bf 1.6\times 10^6}$&\cite{Tanabashi:2018oca}\\
                          & $B_d \to \mu^+  \mu^-  a$&--&$2.8\times 10^5$&\cite{Albrecht:2019zul}\\
                         & &&$(1.2\times 10^6$)\\
                         &$B-\bar B$ (MFV)& $1.1\times 10^6$& ${\bf 2.0\times 10^6}$&\multirow{2}{*}{\cite{UTfitweb}}\\  
                         &~~~~~~~~~~(gen.)& $1.3\times 10^6$& ${\bf 2.3\times 10^6}$&\\                           
\hline \\[-3mm]

                         $t\to u$ & $K^+\to\pi^+a$ (loop) & $3 \times 10^8~^\dagger$ & $3 \times 10^8~^\dagger$&\multirow{2}{*}{\cite{Adler:2008zza}}\\
$t\to c$ & $K^+\to\pi^+a$ (loop) & $7 \times 10^8~^\dagger$ & $7 \times 10^8~^\dagger$\\
\hline\hline
\end{tabular}
\caption{90\% CL lower bounds on the scales of flavor-violating axion couplings $F_{ij}^V$ and $F_{ij}^A$ (in GeV), with future projections in parentheses. Bounds obtained from data with large experimental or theoretical systematic errors are marked with a  $\dagger$ superscript. The most relevant constraints in each sector are typeset in ${\bf boldface}$.
}
\label{tab:main:results}
\end{table}

In Table~\ref{tab:main:results} we summarize the 90\% CL lower bounds on the couplings $F_{ij}^{V,A}$, Eq.~(\ref{eq:FVA:def}), obtained using the flavor-changing processes discussed in the previous sections. Some of the bounds are afflicted by large theoretical uncertainties which could in principle change the quoted numerical values by as much as an order of magnitude. The affected bounds are: $(i)$ the bounds on $F_{sd}^{A,V}$ from   supernova cooling due to $\Lambda\to n a$ transition, where the temperature of the PNS and the interpretation of the SN~1987A neutrino events are two important sources of  potential systematic errors; $(ii)$ the meson-mixing bounds from  $\Delta m_D$ and from kaon mixing on $F_{ds}^V$ suffer from poorly-known theoretical predictions;  $(iii)$  and bounds on top-axion couplings rely on additional, model-dependent assumptions on the absence of cancellations with tree-level contributions. In Tab.~\ref{tab:main:results} all these bounds are flagged by a ``$\dagger$'' superscript. Future projections for the bounds based on ongoing or future experiments are given in Tab~\ref{tab:main:results} inside parentheses and will be discussed below. Furthermore, we recall that all meson-mixing bounds are sensitive to additional contributions from UV physics.

\begin{figure*}[t!]
 \includegraphics[width=16cm]{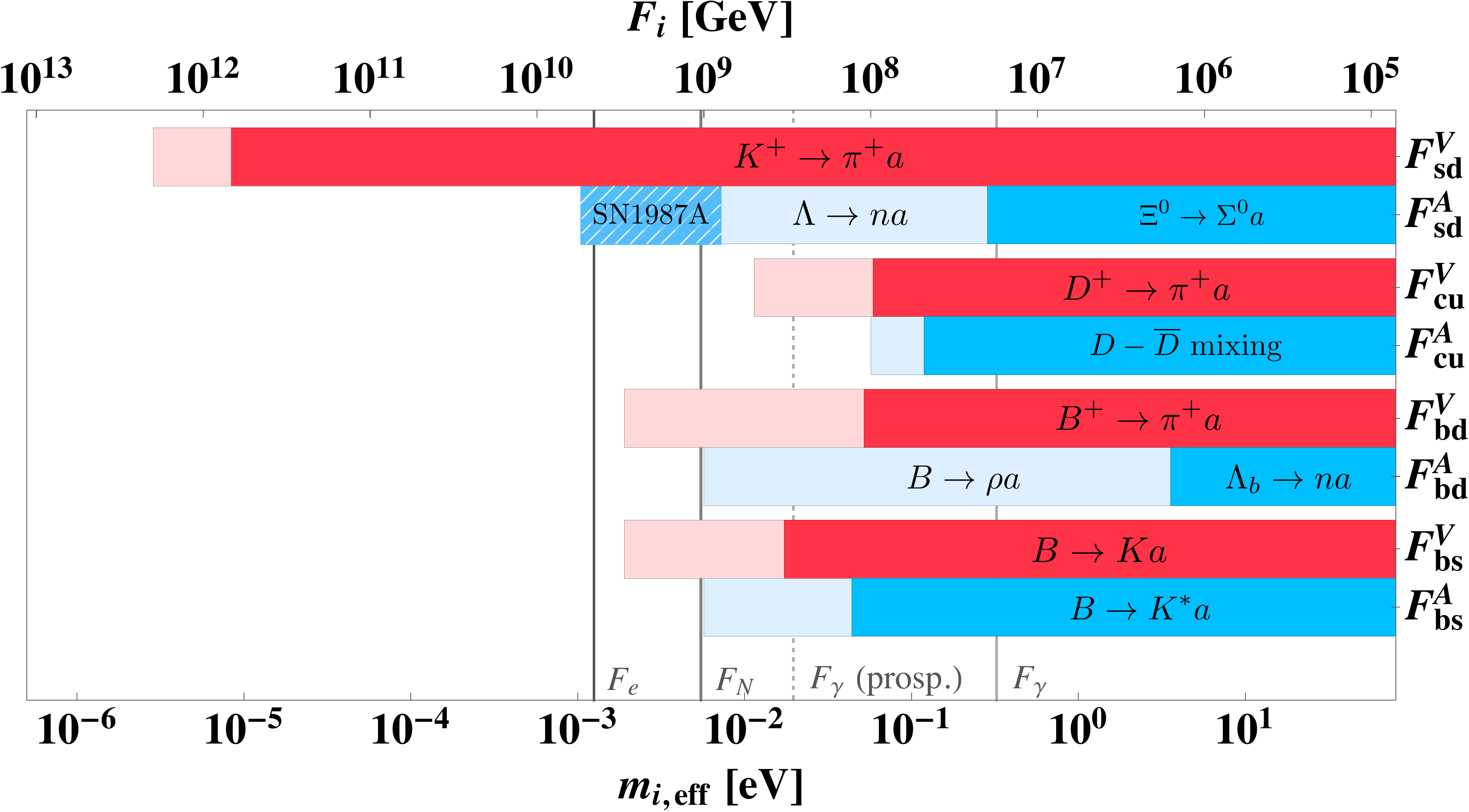} 
\caption{Summary of the most important bounds for the different flavor sectors and for vectorial (red) and axial-vectorial (blue) couplings. On the lower axis we indicate the corresponding values for the effective axion mass defined by $m_{i. {\rm eff}} \equiv 4.69 \eV \times 10^6 \GeV/F_i$. 
 Also shown as vertical gray lines are the bounds on axion couplings to electrons $F_e$ (95\%CL), nucleons $F_N$, and photons $F_\gamma$ (95\%CL), see Section \ref{sec:diag} for details. 
\label{fig:plot}}
\end{figure*}

In Fig.~\ref{fig:plot} we summarize the most relevant bounds for the different flavor sectors and types of couplings, as well as the potential reach of ongoing and future experiments. These are compared to the strongest constraints on the diagonal axion couplings to electrons (WD cooling), nucleons (SN 1987A) and photons (HB/CAST), which were discussed in Section \ref{sec:diag}. For the projected reach on the axion photon coupling ``$F_\gamma$(prosp.)" we quote the prospects for IAXO~\cite{Irastorza:2018dyq, Armengaud:2019uso}. In Fig.~\ref{fig:plot} we do not show the bounds from  ADMX, since these depend heavily on the assumed axion mass, but for particular axion mass ranges can be much stronger than the  bounds from helioscopes.

\subsection{Present bounds}

The strongest bound on QCD axion couplings is of the order of $10^{11-12}$ GeV and due to the stringent constraints on $F^V_{sd}$ from $K^+\to\pi^+ a$ decays, cf. Table~\ref{tab:main:results}. This bound exceeds even the stellar axion bounds, $F_{e,N}\gtrsim 10^{9}$ GeV, which rely on the diagonal couplings. If the flavor structure is not completely generic, the vectorial $sd-a$ couplings could be suppressed  and other probes can become equally or more important (so clearly $K^+\to\pi^+ a$ should not be interpreted as the strongest constraint on the QCD axion independently of the underlying axion model). For instance, as discussed in Sec.~\ref{sec:axion:couplings}, the $sd-a$ couplings could be strongly suppressed by suitable alignment of PQ-charge matrices and SM Yukawas.

The axial-vector $sd-a$  couplings  can be accessed from three-body kaon decay, $K^+\to \pi^+\pi^0 a$, from hyperon decays, and indirectly from $K-\bar K$ mixing, resulting in lower bounds that are in the $\sim {\mathcal O} (10^{6-7})$ GeV range. In principle, the best bound on the axial $sd-a$ coupling, $F_{sd}^A\gtrsim {\mathcal O}(10^{9})$ GeV, comes from hyperon $\Lambda\to na$  transitions in the SN 1987A supernova. In fact, at face value this is the strongest bound of all the axial-vector axion--quark couplings in our analysis. However, as pointed out already above, the supernova bounds should be used with caution due to difficult to estimate systematics. As discussed below, quite impressively the projected improvements on the $\Lambda\to n a$ 
decay branching ratio reach at BESIII will start to compete with this supernova bound, but using well-controlled transitions measured in the lab. 

The two-body decays of $B$ mesons probe all the $b\to s$ and $b\to d$ couplings up to a scale of $\sim 10^8$ GeV, with the exception of $F^A_{bd}$ due to the absence of dedicated $B\to\rho a$ searches  at the $B$ factories (the related  Belle $B\to\rho\nu\bar\nu$ analysis cannot be readily recast, as discussed in Sec.~\ref{sec:bounds:hadron:decays}). The bounds from two-body heavy-baryon $\Lambda_b$ decays and  from $B_{d,s}-\bar B_{ds}$ mixing are more than two orders of magnitude less sensitive. Only the bound on the axial-vector $bd-a$ coupling from  $\Lambda_b$ decays and $B-\bar B$, in the $10^6-10^7$ GeV ballpark, are phenomenologically relevant constraints. Note that in this work we have focused on the case of the QCD axion, such that $b\to q a$ transitions result in the axion escaping the detector and a missing energy signature. A more inclusive search strategy, that does not require any information about the axion decay modes is possible using searches for $B_s\to \mu\mu a$ decays \cite{Albrecht:2019zul}, though with a reduced sensitivity to the quark-axion couplings compared to the other modes. 

The $K$ and $B$ meson decays also probe top-quark axion couplings indirectly through loop effects, cf.~Sec.~\ref{sec:axion:couplings} and Appendix \ref{sec:app:RGEs}. 
These constraints become important in the scenarios with down-quark alignment or flavor-diagonal (or MFV) couplings in the UV. In particular, the strong bound on  the $K^+\to\pi^+ a$ branching fraction translates into bounds for the $t\to c$ and $t\to u$ transitions at the level of $10^8-10^9$ GeV. Although we show these constraints in Table~\ref{tab:main:results}, we did not include them in Fig.~\ref{fig:plot} as they only constrain the left-handed combination and require the absence of possible cancellations with tree-level $sd-a$ couplings.             

Direct probes of flavor violating up-quark--axion couplings that are potentially sensitive to relatively high scales are possible with charmed mesons and baryons. The most sensitive probe of the flavor violating vectorial $cu-a$ coupling turns out to be the two-body $D^+\to \pi^+ a$ decay. 
Recast of the $D^+\to(\tau^+\to\pi^+\nu)\bar\nu$ analysis of CLEO and BESIII gives the bound $F_{cu}^V \gtrsim {\mathcal O}(10^8)$ GeV. 
Axial-vector $cu-a$ couplings are currently probed predominantly by $D^0$-mixing with lower bounds in the range $F_{cu}^A\gtrsim10^6-10^8$ GeV depending on the CP-violating phase of the axion contribution. These couplings can also be directly probed by $\Lambda_c \to p a$ decays, which in the future could provide the best bounds on approximately real $F_{cu}^A$ (this case is not included in Fig.~\ref{fig:plot} for simplicity).  

\subsection{Future projections}

The sensitivity to the couplings of the flavored-axion to the quarks can be greatly improved with future experiments. Dedicated searches for a massless axion in two-body kaon decays at NA62 and KOTO are expected to reach a sensitivity to the branching fraction better than $10^{-11}$~\cite{Ruggieroprivate}  (cf. Sec.~\ref{sec:bounds:hadron:decays}). We thus use 
\begin{align}
{\rm BR_{proj}}(K^+\to\pi^+ a)<10^{-11}, 
\end{align}
as the (conservative) experimental projection.
 As shown in Table~\ref{tab:main:results} this will allow to push the lower bounds on vectorial $sd-a$ couplings beyond a scale of $10^{12}$ GeV.

The three-body kaon decays $K_L\to \pi^0\pi^0 a$ can be potentially searched for at KOTO~\cite{Suzukiprivate}. However, to this date there is no analysis of the sensitivity KOTO could achieve for this decay channel. A direct extension of the current experimental sensitivity for $\text{BR}(K_L\to\pi^0\pi^0 a)\lesssim10^{-6}$~\cite{E391a:2011aa} to the kinematics of a massless axion would give a bound $|F_{sd}^A|\geq6.2\times 10^8$ GeV. A very interesting set of probes for axial $sd-a$ coupling is also offered by the hyperon decays. BESIII has a rich hyperon-physics program, and as part of it searches for axions should be attempted. We use the projections for the $s\to d \nu\bar\nu$ decay modes, estimated for 5 fb$^1$ integrated luminosity in~\cite{Li:2016tlt}, as the projected bounds on the $s\to d a$ decays:
\beq
\begin{split}\label{eq:projhyps}
&{\rm BR_{proj}}(\Lambda\to na)<3\times10^{-7},\\
&{\rm BR_{proj}}(\Sigma^+\to pa)<4\times10^{-7},\\
&{\rm BR_{proj}}(\Xi^0\to \Sigma^0a)<9\times10^{-7},\\
&{\rm BR_{proj}}(\Xi^0\to \Lambda a)<8\times10^{-7}.
\end{split}
\eeq
This will allow to reach scales as high as $|F_{sd}^A|\sim 10^9 \GeV$, entering in the range constrained indirectly by the supernova emissivity bound. Dedicated searches for such $s\to d a$ hyperon transitions could lead to even more stringent constraints than the above conservative estimates. 

All the limits on heavy-quark transitions from two-body meson decays reported in this paper are obtained recasting the analyses of ``$\nu\bar\nu$'' decays at  BaBar and CLEO. These bounds will  certainly be improved by dedicated searches in the future. Belle II expects to gather 50 ab$^{-1}$ of data in the next five years, roughly 
a factor 100 larger than the final integrated luminosity at BaBar.
 The gain in the bounds on the branching ratios depends on the scaling behaviors of the backgrounds. In the absence of dedicated experimental projections, we simply assume an ``optimistic" scaling inversely proportional to the increase in luminosity with respect to the integrated luminosities on which the respective BaBar analyses were performed. The ``conservative" scaling inversely proportional to the square-root of the number of total events  would result in slightly weaker bounds. Assuming similar reconstruction efficiencies at Belle II as those achieved at BaBar 
one can expect an improvement in the sensitivity to $F_{bs}^{V,A}$ and $F_{bd}^{V}$ by at least an order of magnitude, see Table~\ref{tab:main:results} for the optimistic projections. For the conservative scaling the expected bounds are about a factor 5 weaker. 

In case of $B\to\rho a$ the future projection can be estimated by using the current Belle bound for the ``$\nu\bar\nu$'' mode in Table~\ref{tab:ExpInput} and rescaling with the luminosities. This gives us,
\begin{align}
 {\rm BR(B^+\to\rho^+a)_{prosp}}< 4 \times10^{-7},
\end{align}
which would give the sensitivity to $f_a\gtrsim 10^9$ GeV from the axial $bd-a$ couplings, about three orders of magnitude stronger than the current bound on this coupling from $B$-$\bar B$ mixing. The expected bound in case of conservative scaling is about a factor of 3 smaller.    

Finally, the searches for axions in charm-meson and charm-baryon decays could be undertaken at BESIII and Belle II. 
For mesons one can rescale the CLEO bound on $D \to \pi a$ by the expected gain in luminosity which is about a factor 20, assuming that BESIII will collect 20 fb$^{-1}$, which leads to bound on $F^V_{cu}$ that is stronger by a  factor 5 (2) for the ``optimistic" and ``conservative" scalings, respectively.  For baryon decays, $\sim 10^4$ $\Lambda_c^+\bar \Lambda_c^-$ pairs have been produced with $567$ pb$^{-1}$ at BESIII~\cite{Ablikim:2015prg} while  $\sim10^7$ $\Sigma^+\bar\Sigma^-$ pairs are expected with 5 fb$^{-1}$~\cite{Li:2016tlt}. BESIII could therefore reach the limit
\begin{align}
{\rm BR_{proj}}(\Lambda_c\to p a)<4\times10^{-5},
\end{align}
obtained by naively rescaling with the relative sizes of the samples the projection for the branching fraction of the $\Sigma^+\to p a$ decay  shown in Eq.~(\ref{eq:projhyps}). This should lead to bounds on the axial $cu-a$ coupling that are comparable with the current bounds from $D-\bar D$ mixing, but with the benefit of no UV model dependence.


\section{Summary and Conclusions} 
\label{sec:conclusions}

In conclusion, in this paper we explored the phenomenological implications of the possibility that the QCD axion has flavor violating couplings to the SM quarks. Although the presence of flavor violation in axion models is very model-dependent, this scenario generically arises whenever the $U(1)_{\rm PQ}$ charges are not family universal. The flavor violating couplings may even be a necessary ingredient in the UV structure of the theory. This is the case, for instance, if the $U(1)_{\rm PQ}$ group that solves the strong CP problem is part of a larger flavor symmetry group, which is a frequent feature in models addressing the SM flavor problem with approximate horizontal symmetries. 

In this paper we investigated in detail the flavor phenomenology of the axion couplings to quarks. We advocate to use a model-independent approach, treating every flavor-violating coupling, vectorial or axial-vectorial, as a different parameter. This is very similar, in spirit, to the global analyses to search for heavy new physics in low-energy experiments and the LHC using the SM effective field theory. Throughout this work we have assumed the limit of a practically massless axion. Much of the framework developed in this paper can be extended straightforwardly to massive axion-like particles with the appropriate changes due to the different kinematics involved.  

The present paper goes beyond previous studies in the same spirit~\cite{Feng:1997tn, Kamenik:2011vy, Bjorkeroth:2018dzu} in several ways: \\
\indent
{\bf \textit{(i)}} We critically examined the bounds that can be derived using two-body decays of heavy-mesons with final state axions. By recasting BaBar data of $B\to K^{(*)}\nu\bar\nu$ and $B\to\pi\nu\bar\nu$ one can derive limits that supersede the old CLEO direct searches of $B\to K a$ and $B\to\pi a$. We also find that the Belle analyses on the ``$\nu\bar\nu$'' modes cannot be recast for this purpose. Thus, there is no bound from two-body decays on $F_{bd}^A$, with the best limit currently given by $\Lambda_b \to na $ decays and $B-\bar B$ mixing.    
\\
\indent
{\bf \textit{(ii)}}  We derived the strongest direct limit in the up-quark sector by recasting data on $D^+\to(\tau^+\to\pi^+\bar\nu)\nu$ as a search for the $D^+\to \pi^+ a$ decay.  
\\
\indent
{\bf \textit{(iii)}}  We provided the theoretical framework and phenomenological analysis of new processes not considered before, such as the three-body $K\to\pi\pi a$ decays and baryon decays. We argue that baryon decays can in the future give   the best sensitivity to several couplings of the axion: the CP-conserving $cu-a$ couplings from $\Lambda_c\to p a$ decay, or the axial vector $sd-a$ coupling from hyperon decays.
\\
\indent
{\bf \textit{(iv)}}  We derived the strongest limit on axial couplings using a novel interpretation of the supernova SN~1987A bound, based on the impact of the process $\Lambda\to n a$ on the neutrino emissivity of the hot proto-neutron star.
\\
\indent
{\bf \textit{(v)}}  We  developed a theoretical framework to extract reliable limits from neutral-meson mixing. This is based on the effective field theories of QCD; chiral-perturbation theory for the case of kaon-mixing and operator product expansion for the case of heavy-meson mixing.  
 
Our main results for present and future constraints on flavor-violating axion couplings are summarized in Table \ref{tab:main:results} and Fig.~\ref{fig:plot}. For several of the  considered modes a significant improvement is expected. Precision flavor facilities can potentially test PQ breaking scales up to the order of $10^{12}$ GeV (NA62 and KOTO) and $10^9$ GeV (Belle II and BES III), if dedicated searches are performed in ongoing experiments at strange, charm and bottom factories. This reach actually falls into the most interesting region of the parameter space where the axion can account for the observed dark matter  abundance or possibly explain various mild hints for anomalous stellar cooling. These expectations strongly motivate a comprehensive experimental program of searching for axions in rare flavor-changing transitions, which may well lead to the first discovery of the QCD axion.

\section*{Acknowledgments}

We thank Daniel Craik, Marko Gersabeck, Pablo Goldenzweig, Aida El-Khadra, Martin Heeck, Phil Ilten, Alex Kagan, Hajime Muramatsu, Uli Nierste, Steven Robertson, Luca Silvestrini, Abner Soffer, Sheldon Stone, Olcyr Sumensari, Shiro Suzuki, Giuseppe Ruggiero and Emmanuel Stamou for useful discussions.
JMC acknowledges support from the Spanish MINECO through the ``Ram\'on y Cajal'' program RYC-2016-20672 and the grant  PGC2018-102016-A-I00. JZ acknowledges support in part by the DOE grant DE-SC0011784. PNHV acknowledges support from the Non-Member State CERN Summer Student Programme. 

\appendix

\section{Renormalization Group Equations}
\label{sec:app:RGEs}
Above the electroweak scale the Lagrangian describing the interactions of the axion with the SM Higgs and the fermions take the form
\begin{align}
{\cal L}_{\rm UV} & \supset i \sum_\psi  \overline{\psi} \gamma^\mu D_\mu \psi + (D^\mu H)^\dagger D_\mu H \nonumber \\
& +  y^u_{ij} H \overline{q}_L^i  u_R^j +  y^d_{ij} \tilde{H} \overline{q}_L^i  d_R^j + y^e_{ij} \tilde{H} \overline{\ell}_L^i  e_R^j + {\rm h.c.} 
\nonumber \\
& + c^q_{ij} \frac{\partial_\mu a}{2 f_a} \overline{q}_L^i \gamma^\mu  q_L^j + c^u_{ij} \frac{\partial_\mu a}{2 f_a} \overline{u}_R^i \gamma^\mu  u^j_R \nonumber \\
& + c^d_{ij} \frac{\partial_\mu a}{2 f_a} \overline{d}_R^i \gamma^\mu  d_R^j + c^\ell_{ij} \frac{\partial_\mu a}{2 f_a} \overline{\ell}_L^i \gamma^\mu  \ell^j_L \nonumber \\
& + c^e_{ij} \frac{\partial_\mu a}{2 f_a} \overline{e}_R^i \gamma^\mu  e^j_R + c_H \frac{\partial_\mu a}{2 f_a} i H^\dagger \overset{\leftrightarrow}{D^\mu} H \, ,
\end{align} 
where $H^\dagger \overset{\leftrightarrow}{D^\mu} H =H^\dagger {D^\mu} H- (D_\mu H)^\dagger  H$. In the first two lines we included the relevant parts of the SM Lagrangian -- the fermion kinetic terms, the Higgs kinetic term and the Yukawa interactions. For axion interactions we keep only the lowest dimension operators and allow for general flavor structure. The Yukawa matrices $y_{ij}^f$ are general $3\times 3$ matrices, while the axion interactions are described by hermitian $3\times 3$ matrices $c_{ij}^f$.   

The Yukawa interactions are invariant under the axion-dependent, flavor-diagonal field redefinitions
\begin{align}
\label{eq:axion:transf}
H & \to e^{  i \alpha \frac{a(x)}{2 f_a}} H \, ,  & \psi_i & \to e^{ - i \alpha Y_\psi  \frac{ a(x)}{f_a}} \psi_i \, ,
\end{align}
where $\alpha$ is a free parameter and $Y_\psi$ denotes hypercharge. Under these transformations the axion couplings shift as 
\begin{align}
\label{eq:axion:shifts}
c_H & \to c_H - \alpha \, , & c^\psi_{ij} & \to c^\psi_{ij} + 2 \alpha Y_\psi \delta_{ij} \, .
\end{align}
Thus one can always employ the above field redefinition with $\alpha = c_H$ to get rid of the $c_H$ operator, and shift the axion couplings to fermions as in Eq.~\eqref{eq:axion:shifts}, which gives Eq.~\eqref{eq:couplings}.
The axion couplings to the gauge fields also do not change, since the transformations in \eqref{eq:axion:transf} are non-anomalous --
 they correspond to a phase shift of each fermion field that is proportional to its 
 hypercharge  $Y_\psi$. When performing the renormalization group (RG) evolution this field redefinition needs to be performed at each scale $\mu$, in order to keep $c_H(\mu)=0$.

The RG equations for the couplings in matrix notation, keeping $c_H\ne0$, are given by\footnote{One can verify that these equations transform consistently under the field redefinitions in Eq.~\eqref{eq:axion:shifts}.}  (see,  e.g., Ref.~\cite{Choi:2017gpf}),
\begin{align}
\begin{split}
16 \pi^2 \frac{d {\bf c}_q}{d \ln \mu} & =  \frac{1}{2} \left( {\bf y}_u {\bf y}_u^\dagger +  {\bf y}_d {\bf y}_d^\dagger\right) {\bf c}_q - {\bf y}_u {\bf c}_u {\bf y}_u^\dagger   \\
 + &   \frac{1}{2} \, {\bf c}_q \left( {\bf y}_u {\bf y}_u^\dagger +  {\bf y}_d {\bf y}_d^\dagger\right) -  {\bf y}_d  {\bf c}_d {\bf y}_d^\dagger  
\\
 - &   c_H \left( {\bf y}_u {\bf y}_u^\dagger -  {\bf y}_d {\bf y}_d^\dagger \right)  \, , 
\end{split}
\\
\begin{split}
16 \pi^2 \frac{d {\bf c}_u}{d \ln \mu} & = {\bf c}_u {\bf y}_u^\dagger {\bf y}_u +  {\bf y}_u^\dagger {\bf y}_u {\bf c}_u -  2 \, {\bf y}_u^\dagger{\bf c}_q  {\bf y}_u    \\
 + &  2 \, c_H  {\bf y}_u^\dagger {\bf y}_u   \, , 
\end{split}
\\
\begin{split}
16 \pi^2 \frac{d {\bf c}_d}{d \ln \mu} & = {\bf c}_d {\bf y}_d^\dagger {\bf y}_d +  {\bf y}_d^\dagger {\bf y}_d {\bf c}_d -  2 \, {\bf y}_d^\dagger{\bf c}_q  {\bf y}_d  \\
- &  2 \, c_H  {\bf y}_d^\dagger {\bf y}_d \, , 
\end{split}
\\
\begin{split}
16 \pi^2 \frac{d {\bf c}_\ell}{d \ln \mu} & =  \frac{1}{2}  {\bf y}_e {\bf y}_e^\dagger  {\bf c}_\ell +   \frac{1}{2} \, {\bf c}_\ell  {\bf y}_e {\bf y}_e^\dagger  - {\bf y}_e {\bf c}_e {\bf y}_e^\dagger    \\
 + &   c_H   {\bf y}_e {\bf y}_e^\dagger \, , 
\end{split}
\\
\begin{split}
16 \pi^2 \frac{d {\bf c}_e}{d \ln \mu} & = {\bf c}_e {\bf y}_e^\dagger {\bf y}_e +  {\bf y}_e^\dagger {\bf y}_e {\bf c}_e -  2 \, {\bf y}_e^\dagger{\bf c}_\ell  {\bf y}_e   \\
 - &  2 \, c_H  {\bf y}_e^\dagger {\bf y}_e   \, , 
\end{split}
\\
\begin{split}
16 \pi^2 \frac{d c_H}{d \ln \mu} & = 6 \,  {\rm Tr} \left( {\bf c}_q {\bf y}_u {\bf y}_u^\dagger  - {\bf c}_q {\bf y}_d {\bf y}_d^\dagger   \right)  \\
 +& 6 \, {\rm Tr} \left(  {\bf c}_d  {\bf y}_d^\dagger {\bf y}_d   -  {\bf c}_u  {\bf y}_u^\dagger {\bf y}_u   \right)  \\
+& 2  \, {\rm Tr} \left(   {\bf c}_e  {\bf y}_e^\dagger {\bf y}_e - {\bf c}_\ell {\bf y}_e {\bf y}_e^\dagger   \right)  \\
- & 2 \, c_H   {\rm Tr} \left( 3 {\bf y}_u {\bf y}_u^\dagger  +  3 {\bf y}_d {\bf y}_d^\dagger + {\bf y}_e {\bf y}_e^\dagger \right)   \, .
\end{split}
\end{align}
Using these equations, one can express the low-energy couplings in terms of high-energy couplings, diagonal SM Yukawas and the CKM matrix $V$. Keeping only the effects proportional to the top yukawa coupling, the radiative corrections to the axion couplings, $\Delta c^{V,A}_{f_i f_j}$, are given by
\begin{align}
\begin{split}
\label{DeltaytA}
\Delta c^{V}_{d_i d_j} (\mu) & =   \frac{y_t^2}{64 \pi^2} \log\frac{f_a}{\mu}  \Big[ 2 V^*_{3i} V_{3j} (c^V_{ tt} + c^A_{tt} )  
\\
&-  V^*_{ki} V_{3j}   (c^V_{u_k t} - c^A_{u_k t} )    \\
& -    V^*_{3i} V_{kj} (c^V_{t u_k} - c^A_{t u_k} )   \Big] \, ,
\end{split}
 \\
 \begin{split}
\Delta c^{A}_{d_i d_j} (\mu) & =  - \frac{y_t^2}{64 \pi^2} \log\frac{f_a}{\mu}  \Big[  24 \, c^A_{tt} \delta_{ij} \\
&+  2 V^*_{3i} V_{3j} (c^V_{ tt} + c^A_{tt} )  \\
&-  V^*_{ki} V_{3j}   (c^V_{u_k t} - c^A_{u_k t} ) \\
&-   V^*_{3i} V_{kj} (c^V_{t u_k} - c^A_{t u_k} )     \Big] \, , 
\end{split}
\\
 \begin{split}
\Delta c^V_{u_i u_j} (\mu) & =   \frac{y_t^2}{64 \pi^2} \log\frac{f_a}{\mu}  \Big[   2 (3 c^V_{tt} - c^A_{tt} )  \delta_{it}  \delta_{jt}  \\
&-  \big(3 c^V_{t u_j} + c^A_{t u_j} \big) \delta_{it} \\
&- (3 c^V_{u_i t} + c^A_{u_i t} ) \delta_{jt}  \Big] \, ,
 \end{split}
 \\
  \begin{split}
\Delta c^A_{u_i u_j} (\mu) & =   \frac{y_t^2}{64 \pi^2} \log\frac{f_a}{\mu}  \Big[  24 \, c^A_{tt} \delta_{ij}  \\
&-    (c^V_{t u_j} + 3 c^A_{t u_j} ) \delta_{it} \\
&- ( c^V_{u_i t} + 3 c^A_{u_i t} ) \delta_{jt}   \\
&+  2 ( c^V_{tt} - 3 c^A_{tt} )  \delta_{it}  \delta_{jt}  \Big] \, , 
\end{split}\\
\Delta c^A_{e_i e_j} (\mu) & =   - \frac{6 y_t^2}{16 \pi^2} \log\frac{f_a}{\mu}   c^A_{tt} \delta_{ij} \, .
\label{DeltaytB}
\end{align}
The flavor-universal contributions arise from a non-zero $c_H$ which is radiatively generated and then through field redefinitions absorbed into $c_{f_i f_j}$ at low energies. 
Note also that the high-energy couplings satisfy 
\begin{align}
c^V_{u_i u_j} -  c^A_{u_i u_j} = V_{ik} V^*_{jl} \left[ c^V_{d_k d_l} -  c^A_{d_k d_l} \right]  \, . 
\end{align}

\section{Details on two-body recasts }
\label{sec:app:recast}
For the recasts of $P_1\to P_2 \nu\bar \nu$ and $P_1\to V_2 \nu\bar \nu$ we use the experimental information  
in the kinematic regions corresponding to a massless axion, i.e., taking into account only the bin that contains events with vanishing invariant mass of the neutrino pair~\footnote{Since usually the bin size is wider than the experimental momentum resolution, in this way we count also background or SM events with $m_{\nu \overline{\nu}} \ne 0$ as axion signals. This renders the resulting bound only more conservative, which will be eventually superseded by a proper recast done by the experimental collaborations using the full information.}. For $B \to \pi a$ we take from the BaBar analysis in Ref.~\cite{Aubert:2004ws} for the numbers of observed and background events in the relevant bin $N_{\rm obs} = 1$ and $N_{\rm bg} = 1$, respectively, for the total number of $B$-mesons in the data sample $N_{\rm tot} = (8.9 \pm 0.1) \times 10^7$, and for the efficiency $\eps = (6.5 \pm 0.6) \times 10^{-4}$. The expected number of events is then $\mu = N_{\rm bg} +  \eps \,  N_{\rm tot} \,{\rm BR} (B \to \pi a)$. To obtain the 90\% CL upper limits we follow the statistical treatment in Ref.~\cite{BABAR13} and use the mixed frequentist-Bayesian approach described in Refs.~\cite{Barlow:2002bk,Cousins:1991qz} in order to include systematic uncertainties. 

In the same way we proceed for  $B \to K a$ and  $B \to K^* a$ using the BaBar data from Ref.~\cite{Adler:2008zza} for the two decay channels in each case, and  $N_{\rm tot} = (4.71 \pm 0.03 )\times 10^8$. For the channels $\{B^+ \to K^+  \nu \overline{\nu},B^0 \to K^0 \overline{\nu} \nu \} $ we take $N_{\rm obs} = \{ 2, 0\}$, $N_{\rm bg} = \{ 0,0 \}$ and  $\eps = \{ 9.5 \pm 0.5, 4.5 \pm 0.5 \} \times 10^{-4}$, while for  $\{B^+ \to K^{*+}  \nu \overline{\nu},B^0 \to K^{*0} \overline{\nu} \nu \} $ we use $N_{\rm obs} = \{ 1, 3\}$, $N_{\rm bg} = \{ 1,1 \},  \eps = \{ 1.2 \pm 0.1, 0.3 \pm 0.05 \} \times 10^{-4}$. In each case the two channels are combined by maximizing the likelihood function that is the product of Poisson probabilities, following Ref.~\cite{Barlow:2002bk}. 

Finally, for $c\to u$ transitions we use the search for $D^+ \to \tau^+ \nu$ decay in the $\tau^+\to \pi^+\bar \nu$ channel at CLEO in Ref.~\cite{CLEO2008}. In the signal window $m_{\rm miss}^2\leq 0.05{\rm~GeV}^2$ CLEO observed  $N_{\rm obs} = 11$ pion like events, with the expected SM background of $N_{\rm bg} = 13.5 \pm 1.0$, for a total number of tagged $D$ decays $N_{\rm tot} = 4.6 \times 10^5$ and single pion detection efficiency $\eps_\pi = 0.89$. This results in the 90\%CL upper bound ${\rm BR}(D\to \pi a)\leq 8.0\cdot 10^{-6}$, following the same statistical prescription as before. A recast of the recent $D^+ \to \tau^+ \nu$ BESIII analysis~\cite{Ablikim:2019rpl} using pion like events from the $\tau^+\to \pi^+\bar \nu$ channel, Fig. 3 right in Ref.~\cite{Ablikim:2019rpl}, results in a bound that is about a factor $2$ weaker than our recast of the bound from  CLEO.

\section{Hadronic matrix elements}
\label{sec:app:hadronic}
In this appendix we give further details on the hadronic elements describing axion induced flavor changing transitions. 
The numerical values of different inputs entering the  predictions are collected in Table~\ref{tab:inputs}.

\begin{table*}[t]
\vspace{0.2cm}
\begin{tabular}{ccccc}
\hline\hline
Flavors & Process&\multicolumn{2}{c}{Inputs} &References\\
\hline\\ [-2.ex] 
\multirow{10}{*}{\hspace{0.6cm}$s\to d$\hspace{0.6cm}}&$K\to\pi$& \multicolumn{2}{c}{$f_+(0)= 0.9706(27)$}& \cite{FLAG2019}\\
                         &\multirow{2}{*}{$K\to\pi\pi$}&\multicolumn{2}{c}{$f_s=5.705(35)$,~~$f_s^\prime=0.87(5) $,~~$f_s^{\prime\prime}=-0.42(5)$}& \multirow{2}{*}{\cite{Batley:2010zza,Batley:2012rf}}\\
                         &&\multicolumn{2}{c}{$f_p= -0.274(29)$,~~$g_p=4.95(9)$,~~$g_p^\prime=0.51(12)$}&\\
                         \cline{2-4}\\ [-2.ex] 
                         &$B_1\to B_2$ & $f_1(0)$ & $g_1(0)$&\\
                         \cline{2-4}\\ [-2.ex] 
                         &$\Lambda\to n$&$-1.22(6)$ &$-0.89(2)$& \multirow{5}{*}{\cite{Cabibbo:2003cu,Ledwig:2014rfa}}\\
                         &$\Sigma^+\to p$&$-1.00(5)$&$0.34(1)$& \\
                         &$\Xi^-\to \Sigma^-$&$1.00(5)$&$1.26(5)$& \\
                         &$\Xi^0\to \Sigma^0$&$-0.71(4)$&$-0.89(3)$& \\
                         &$\Xi^0\to \Lambda$&$1.22(6)$&$0.24(5)$& \\
                         \cline{2-4} \\ [-2.ex]  
                      &$K-\bar K$ &\multicolumn{2}{c}{$f_{K^\pm}=155.7(3)$ MeV}& \cite{FLAG2019}\\ 
\hline\\ [-2.ex] 
\multirow{3}{*}{$c\to u$}&$D\to\pi$& \multicolumn{2}{c}{$f_+(0)= 0.612(35)$
}& \cite{Lubicz:2017syv}\\
                        &$\Lambda_c\to p$ & \multicolumn{2}{c}{$f_1(0)=0.672(39)$,~~$g_1(0)=0.602(31)$}& \cite{Meinel:2017ggx}\\
                     &$D-\bar D$     &  \multicolumn{2}{c}{$\langle \mathcal O_2\rangle=-0.1442(72),\quad \langle\mathcal O_4\rangle=0.275(14) $}&\cite{Bazavov:2017weg}\\

\hline  \\ [-2.ex]     
\multirow{6}{*}{$b\to s$}&$B\to K$& \multicolumn{2}{c}{$f_+(0)= 0.335(36)$}& \cite{Bailey:2015dka}\\
                          
                         &$B\to K^*$& \multicolumn{2}{c}{$A_0(0)= 0.356(46)$}& \cite{Straub:2015ica}\\
                         &$\Lambda_b\to \Lambda$& \multicolumn{2}{c}{$f_1(0)=0.16(4)$,~~$g_1(0)=0.11(9)$}& \cite{Detmold:2016pkz}\\
                        &\multirow{3}{*}{$B_s-\bar B_s$} & \multicolumn{2}{c}{$f_{B_s}= 230.3(1.3)$ MeV}&\cite{FLAG2019}\\
                        & &  \multicolumn{2}{c}{$B_2=0.817(43),\quad B_4=1.033(47)$}&\cite{Dowdall:2019bea}\\
                         & &  \multicolumn{2}{c}{$\eta_2=-2.669 (62), \quad \eta_4=3.536 (74)$}&\cite{Dowdall:2019bea}\\                       
\hline      \\ [-2.ex] 
\multirow{6}{*}{$b\to d$}& $B\to \pi$&\multicolumn{2}{c}{$f_+(0)= 0.21(7)$}& \cite{Gubernari:2018wyi}\\
            
                        & $B\to \rho$ & \multicolumn{2}{c}{$A_0(0)= 0.356(42)$}& \cite{Straub:2015ica}\\
                         & $\Lambda_b\to p$ & \multicolumn{2}{c}{$f_1(0)=0.23(8)$,~~$g_1(0)=0.12(13)$}& \cite{Detmold:2015aaa}\\
                     &\multirow{3}{*}{$B-\bar B$}    & \multicolumn{2}{c}{$f_{B}=190.0(1.3)$ MeV}& \cite{FLAG2019}\\
                                                &&  \multicolumn{2}{c}{$B_2=0.769(44),\quad B_4=1.077(55)$} & \cite{Dowdall:2019bea}\\
                         & &  \multicolumn{2}{c}{$\eta_2=-2.678 (62), \quad \eta_4=3.547 (74)$}&\cite{Dowdall:2019bea}\\                                                   
\hline\hline                         
\end{tabular}
\caption{Numerical values for the theoretical inputs used in the analysis as described in Appendix~\ref{sec:app:hadronic}. The bag parameters for the $B_{d,s}$ systems are evaluated at the renormalization scale $\mu=m_b$~\cite{Dowdall:2019bea}. The matrix elements of the corresponding four-quark operators $\langle \mathcal O_i\rangle$ in the $D$ meson system are evaluated at $\mu=3$ GeV and are shown in units of GeV$^4$~\cite{Bazavov:2017weg}.  \label{tab:inputs}}
\end{table*}

\subsection{Two-body decays}

We first give the parametrizations of matrix elements for two-body hadron decays $H\to H^\prime a$, where $H^{(')}$ is a pseudoscalar meson, a vector meson or a spin $1/2$ baryon. The transitions are induced by quark level transitions of the type $q\to q^\prime a$. The form factors in the resulting hadronic matrix elements are functions of the momentum exchange squared, $q^2=(p-p^\prime)^2=m_a^2\simeq0$, i.e., for the predictions we only need the values of form factors at $q^2=0$. 

The hadronic matrix elements for $P\to P'a$ transitions, with $P^{(')}$ pseudoscalar mesons, are parametrized by two sets of form factors
\beq
\langle P^\prime\left(p^\prime\right) |\bar q^\prime\gamma^{\mu}q|P\left(p\right)
\rangle= P^\mu f_+^{PP'}(q^2) +q^\mu f_-^{PP'}(q^2),
\label{eq:PtoPFF} 
\eeq
where $P^\mu=(p+p^\prime)^\mu$. The related matrix element of the axial current $\bar q^\prime\gamma^{\mu}\gamma_5 q$ vanishes by parity invariance of the strong interactions.
In the decay the Lorentz index in \eqref{eq:PtoPFF} is contracted with $-iq_\mu$ from the derivative of the axion field, cf. Eq.~(\ref{eq:couplings}). The only hadronic inputs needed to describe the $P\to P'a$ decays are therefore $f_+^{PP'}(0)$. 

For $f_+^{K^+\pi^+}(0)$ we use the $N_f=2+1+1$ lattice QCD determination of $f_+^{K^0\pi^-}(0)$~\cite{FLAG2019}, since in the isospin symmetric limit 
$f_+^{K^+\pi^+}(0)=f_+^{K^0\pi^-}(0)$.
Likewise, we use for the axion induced charm meson decays the lattice QCD calculation of $f_+^{D^0\pi^-}(0)$ reported by the ETM collaboration~\cite{Lubicz:2017syv}, along with the isospin relation
$f_+^{D^+\pi^+}(0)=f_+^{D^0\pi^+}(0)$. For the $B^+ \to K^+ a$ decay we use the lattice QCD determination of the form factors by the Fermilab/MILC collaboration~\cite{Bailey:2015dka}, while for the $B^+ \to \pi^+ a$ decay we use the light-cone sum rule determination of the form factors in~\cite{Gubernari:2018wyi}.

The hadronic matrix element for the decays of a pseudoscalar meson $P$ into a vector meson $V$ is given by
\beq
\begin{split}
\label{eq:PtoVFF}
\langle &V(p^\prime, \eta) | \bar q^\prime \gamma_\mu \gamma_5 q | P(p) \rangle
= i (\eta^* \cdot q) \frac{q_\mu}{q^2} 2\, m_V A_0(q^2) \\
&+ i (m_B+m_V) \left(\eta^{*\mu} - \frac{(\eta^*
  \cdot q) q^\mu}{q^2} \right) A_1(q^2) \\
& - i  (\eta^* \cdot q)\! \left( \frac{(2p-q)^\mu}{m_B\! +\! m_V}
             - (m_B\!-\!m_V) \frac{q^\mu}{q^2} \right) A_2(q^2),
\end{split}
\eeq
where $\eta$ is the polarization vector of the vector meson.
Parity conservation implies that the matrix element of the vector current, $\langle V(p^\prime, \lambda) | \bar q^\prime \gamma_\mu q | P(p) \rangle$, transforms as an axial vector and must be $\propto\epsilon_{\mu\nu\rho\sigma}p^\nu p^{\prime\rho} \eta^\sigma$. This gives a vanishing contribution to the decay amplitude upon contraction with the derivative interaction of the axion. Furthermore, contracting Eq.~(\ref{eq:PtoVFF}) with $-iq^\mu$, and taking into account that
\begin{align}
\frac{m_P+m_V}{2m_V}A_1(0)-\frac{m_P-m_V}{2m_V}A_2(0)=A_0(0), 
\end{align}
one finds that $A_0(0)$ is the only hadronic input needed to describe the $P\to V a$ decays. For the $B\to K^*$ and $B\to\rho$ form factors we use the 
light-cone sum rules determinations from Ref.~\cite{Ball:2004rg}.

The hadronic matrix elements for $B\to B'a$ decay, with $B^{(')}$ the spin-1/2 baryons, are 
parametrized by 
\begin{align}
\label{eq:baryonFFs}
\begin{split}
\langle B^\prime(p^\prime) | &\bar{q}^\prime \gamma_\mu q |B (p)\rangle
=
\bar{u}^\prime (p^\prime)  \Big[
f_1(q^2)  \,  \gamma_\mu  
\\  
&+ \frac{f_2(q^2)}{M}   \, \sigma_{\mu \nu}   q^\nu  
+ \frac{f_3(q^2)}{M}   \,  q_\mu  
\Big]  
 u (p),  
 \end{split}
 \\
 \begin{split}
\langle B^\prime(p^\prime) | &\bar{q}^\prime \gamma_\mu \gamma_5 q |B (p)\rangle
=
\bar{u}^\prime (p^\prime)  \Big[
g_1(q^2)    \gamma_\mu
\\    
&+ \frac{g_{2} (q^2)}{M}   \sigma_{\mu \nu}   q^\nu  
+ 
\frac{g_{3} (q^2)}{M}   q_\mu  
\Big]  \,\gamma_5  u(p).
\end{split}
\end{align}
After contracting with $-i q^\mu$ from the derivative on the axion field, the decay amplitude  involves only two form factors at $q^2=0$, i.e., the vector and axial-vector couplings $f_1(0)$ and $g_1(0)$.
In numerical evaluations
we use for $\Lambda_b \to n$ form factors the lattice QCD determinations of the $\Lambda_b\to p$ form factors from Ref.~\cite{Detmold:2015aaa} (they are equal in 
the isospin-symmetric limit), for $\Lambda_c \to p$ from Ref.~\cite{Meinel:2017ggx}, and for $\Lambda_b\to \Lambda$ from Ref.~\cite{Detmold:2016pkz}. For hyperons we use the flavor SU(3) symmetry to obtain the form factors from their electric charges and semileptonic decays~\cite{Cabibbo:2003cu,Ledwig:2014rfa}. For reference we quote also the uncertainties due to the expected sizes of the SU(3) flavor-breaking effects. The leading corrections to $f_1(0)$ vanish because of the Ademollo-Gatto theorem~\cite{Ademollo:1964sr}. Explicit calculations using lattice QCD showed that the breaking effects are $\lesssim5\%$~\cite{Sasaki:2017jue}. For the axial couplings, $g_1(0)$, the leading SU(3)-flavor breaking corrections can be predicted in chiral perturbation theory by using experimental measurements of the isospin-related channels, the semileptonic hyperon decays, and lattice QCD results~\cite{Ledwig:2014rfa}. 

\subsection{$K\to \pi\pi a$}

The hadronic matrix elements needed for the decays $K^+\to\pi^+\pi^0~a$ and $K_L\to\pi^0\pi^0~a$ can be obtained from the form factors measured in the decay $K^+\to\pi^+\pi^-~e^+\nu$ using isospin symmetry. The matrix element of the axial-vector current for the latter process is defined as~\cite{Pais:1968zza,Bijnens:1994ie},
\begin{align}
\langle &\pi^+(p_+)\pi^-(p_-)|\bar s\gamma^\mu\gamma_5u|K^-(p)\rangle=\nonumber\\
&-\frac{i}{m_K}\Big(F(p_+ + p_-)^\mu+G(p_+-p_-)^\mu+R q^\mu\Big),
\label{eq:KtopipiFF}
\end{align}
where $q=p-p_+-p_-$ 
is the four-momentum of the axion and thus $q^2\simeq 0$. Once the above hadronic matrix element is contracted with $-i q^\mu$ from the derivative on the axion field, the contribution of $R$ to the decay matrix element 
vanishes at the physical point $q^2=0$. The related matrix element of the vector current completely vanishes due to parity invariance once it is contracted with the axion-field derivative, similarly to the  $P\to V a$ decays discussed above.  

The $K\to \pi\pi$ form factors depend on three kinematic variables which can be chosen to be $q^2$, $s=(p_++p_-)^2$ and $\cos\theta_\pi$, where $\theta_\pi$ is the angle between the positively-charged pion's and kaon's three-momenta in the di-pion rest frame. The form factors are complex functions with a strong phase arising due to the rescatterings of the pions. We follow a standard parametrization of the form factors using a partial wave expansion of the two-pion system~\cite{Pais:1968zza,Bijnens:1994ie}, and truncate it at the $p$-wave,
\begin{align}
F&=F_s+F_p\cos\theta_\pi\exp(-i\delta),\\
G&=G_p\exp(-i\delta).
\end{align}
Here  $\delta=\delta_s-\delta_p$ is the difference of $s$- and $p$-wave $\pi^+\pi^-$ phase shifts. The prefactors $F_s$, $F_p$ and $G_p$ are only functions of $q^2$ and $s$. In the experimental analyses of $K\to \pi\pi e\nu$ decays they are often Taylor expanded around $q^2=0$ and $s=4 m_\pi^2$. For the $K\to \pi\pi a$ decays we only need their values at $q^2=0$, retaining the parameters of the expansion in the dimensionless variable $\bar s=(s/4m_\pi^2)-1$, around $\bar s=0$, 
\begin{align}
F_s&=f_s+f_s^\prime~\bar s+f_s^{\prime\prime}~\bar s^2, \nonumber\\
F_p&=f_p,\quad G_p=g_p+g_p^\prime~\bar s,
\end{align}
which are determined from the experimental data~\cite{Batley:2010zza,Batley:2012rf} and shown in Table~\ref{tab:inputs}. 

In order to connect these form factors in the  $K^+\to\pi^+\pi^-e^+\nu$ channel to those in $K^+\to\pi^+\pi^0~a$ and $K_L\to\pi^0\pi^0~a$, one uses the isospin-symmetry relations~\cite{Littenberg:1995zy,Chiang:2000bg} with the convention that $(u,~d)$ and $(-\bar d,~\bar u)$ transform as isodoublets, 
\begin{align}
\begin{split}
    \langle \pi^+\pi^0|\bar s\gamma^\mu&\gamma_5d|K^+\rangle=
    \\
    &-\sqrt{2}\langle (\pi^+\pi^-)_{I=1}|\bar s\gamma^\mu\gamma_5u|K^+\rangle,
    \end{split}
    \\
    \begin{split}
    \langle \pi^0\pi^0|\bar s\gamma^\mu&\gamma_5d|K^0\rangle=
    \\
   & \langle (\pi^+\pi^-)_{I=0}|\bar s\gamma^\mu\gamma_5u|K^+\rangle, 
    \end{split}
\end{align}
where the subscripts on the right-hand sides indicate that we have projected the isospin wave functions of the two final pions to either $I=0$ or $I=1$. The total amplitude must be even under the exchange of the two pions (Bose symmetry). Therefore, for the isospin symmetric $I=0$ (anti-symmetric $I=1$) wave function only the $s$-wave ($p$-wave) components contribute. 
A final observation is that one does not have interference in the total rates between $s$- and $p$-wave components in these axion decay channels and they are insensitive to the strong phase $\delta$. 

\subsection{Neutral meson mixing}

Hadronic matrix elements of the four-quark operators Eq.~(\ref{eq:local_ops}) involved in the axion contributions to heavy neutral-meson mixing are conventionally defined in terms of the so-called bag parameters, $B_i$. For the case of $B-\bar B$ mixing and shortening $\langle B^0|\mathcal O_i|\bar B^0\rangle=\langle \mathcal O_i\rangle$ these are defined as,
\beq
\langle B_q^0|O_i^q|\bar B_q^0 \rangle=\frac{1}{4}\eta_i^q(\mu) f_{B_q}^2 m_{B_q}^2 B_{B_q}^{(i)}(\mu),
\eeq
with the values for $\eta_i^q(m_b)$ and $B_{B_q}^{(i)}(m_b)$ as provided in~\cite{Dowdall:2019bea}.
These definitions are straightforwardly extended to the short-distance contributions in the other neutral-meson systems. The values of the different parameters in these equations are obtained from lattice calculations. In case of the charm-meson oscillations we use results in ref.~\cite{Bazavov:2017weg} which directly provides the results in terms of the matrix elements $\langle \mathcal O_i\rangle$ at $\mu=3$ GeV.

\bibliographystyle{apsrev4-1.bst}
\bibliography{refs}

\end{document}